\documentclass[preprint,showpacs,preprintnumbers,amsmath,amssymb,superscriptaddress]{revtex4}

\usepackage{graphicx,color,epsfig}
\usepackage{amsmath,amssymb}
\usepackage{url}
\usepackage{bm}
\usepackage{dcolumn}
\newcommand{\hs}{\hspace*{0.5cm}}

\newcommand{\be}{\begin{equation}}
\newcommand{\ee}{\end{equation}}
\newcommand{\bea}{\begin{eqnarray}}
\newcommand{\eea}{\end{eqnarray}}
\newcommand{\nn}{\nonumber}
\newcommand{\crn}{\nonumber \\}

\newcommand{\bc}{\begin{center}}
\newcommand{\ec}{\end{center}}

\newcommand {\ba}{\begin{array}}
\newcommand {\ea}{\end{array}}
\newcommand{\ben}{\begin{enumerate}}
\newcommand{\een}{\end{enumerate}}

\begin{document}

\title{ Signal of doubly charged Higgs at $e^+e^-$ colliders}

\title{ Signal of doubly charged Higgs at $e^+e^-$ colliders}

\author{L. T. Hue}
\affiliation{Institute of Physics,   Vietnam Academy of Science and Technology, 10 Dao Tan, Ba
Dinh, Hanoi, Vietnam\email{lthue@iop.vast.ac.vn} }
\author{D.T. Huong}
\affiliation{Institute of Physics,   Vietnam Academy of Science and Technology, 10 Dao Tan, Ba
Dinh, Hanoi, Vietnam }
\author{H.N. Long}
\affiliation{Institute of Physics,   Vietnam Academy of Science and Technology, 10 Dao Tan, Ba
Dinh, Hanoi, Vietnam }
\author{H.T. Hung}
\affiliation{Department of Physics, Hanoi University of Education 2, Phuc Yen, Vinh Phuc, Vietnam}
\author{N.H. Thao\thanks{These authors contributed equally to this work} }
\affiliation{Department of Physics, Hanoi University of Education 2, Phuc Yen, Vinh Phuc, Vietnam}

\begin{abstract}%
Masses and signals of  the production of Doubly charged Higgses
(DCH) in the  framework of the supersymmetric reduced minimal
3-3-1 model (SUSYRM331) are investigated.  In the DCH sector, we
prove that   there always exists a region of the parameter  space
where the mass of the lightest DCH is in  order of $\mathcal{O}(100)$ GeV
  even when all other new particles are very heavy.
  The lightest DCH mainly decays to two same-sign leptons while
   the dominated decay channels of the heavy DCHs are those decay to heavy particles.
    We analyze each
    production cross section for $e^+e^- \rightarrow H^{++} H^{--}$ as a function of
      a few kinematic
    variables, which are useful to discuss  the creation of  the DCHs in
    the $e^+e^-$ colliders  as a signal of
    new physics beyond the Standard Model. The numerical study shows that the cross sections for creating
    the lightest DCH can reach  values of few pbs. The two other DCHs are too
    heavy, beyond the observable range of experiments.
    The lightest DCH may be detected by the International Linear Collider (ILC) or
   the Compact LInear Collider (CLIC) by searching
    its decay  to a same-sign charged lepton pair.

\end{abstract}
\pacs{11.15.Ex, 12.60.Fr, 13.66.Fg }

\maketitle

\section{Introduction}
\label{sec:intro} The detection of  the Higgs boson with mass around 125 GeV  by experiments at  the large hadron
collider (LHC)
\cite{higgsdicovery1,higgsdicovery2,h2ga1,h2ga2} has again confirmed the success of the Standard Model (SM).
However, this model needs to be extended to cover other problems which cannot be explained in this framework, specially
small neutrino mass and mixing, dark matter, asymmetry of matter and antimatter...  Theories that lie beyond the SM not only  solve the SM problems but also predict the signals of new physics which can be searched in the future.  Many well-known  models beyond
the SM   have Higgs spectrum containing the DCHs, for example
 the left-right model \cite{Mohopatra1,Mohopatra2,Mohopatra3}, the
Zee-Babu model \cite{Zee1,Zee2}, the 3-3-1 models \cite{pisano1,rfoot,ffv1,orminimal},... and their supersymmetric versions
\cite{leftrightm1,leftrightm2,leftrightm3,leftrightm4,leftrightm5,sMontero,huong1,Ferr}. The appearance of the DCHs will really
 be one of the  signals of new physics. Hence, there have been numbers of publications predicting this  signal in colliders such as the  LHC,
ILC \cite{ILC1,ILC2} and CLIC \cite{CLIC1,CLIC2}. Recent experimental
searches for the DCHs have been doing at   the   LHC
\cite{H2Ex,H2Ex2,H2Ex3,beringer},  through  the decays of them into a pair of same-sign charged leptons. This decay channel has been investigated in many above models:  the left-right symmetric model \cite{h2lr,h2lr1} and  the supersymmetric version \cite{sh2lr}, the 3-3-1
models \cite{h2331m,h2331m2}. On the other hand, some other SM extensions including Higgs triplets were shown that the DCHs may have main decay channels  of $H^{\pm\pm}\rightarrow W^{\pm}W^{\pm}$ \cite{hww1,hww2}, or $H^{\pm}W^{\pm}$\cite{hwh1}, leading to the lower bounds of DCH masses than those concerned from searching the DCH decay into leptons.  It is noted that the Higgs sectors in the supersymmetric (SUSY) models
 seem to be very interesting
because they do not contain unknown self- couplings of four Higgses in the superpotential,
unlike the case of non-SUSY models where this kind of couplings  directly contribute to Higgs masses. As a consequence,
 some Higgses will get masses mainly from the D-term, namely from the electroweak breaking scale,
  leading to values
 of Higgs masses being in order of $\mathcal{O}(100)$ GeV at the tree level.  This happens in SUSY models such
 as the MSSM, supersymmetric versions of the economical 3-3-1 (SUSYE331) and reduced minimal 3-3-1
 (SUSYRM331) models  \cite{huong1,Ferr,susye331r}.  It was shown that there is at least one neutral
  CP-even Higgs inheriting a tree-level mass below the mass of $Z$ boson, $m_Z=92$ GeV. Fortunately,
   the loop-correction contributions increase the full mass of this Higgs up to the recent experimental value.
   This  suggests that some other Higgses may be light with masses order of  $\mathcal{O}(100)$ GeV.
   In the MSSM, this  cannot happen if soft parameters such as $b_{\mu}$
term relating with mass of the neutral  CP-odd Higgs
    are large.  Other SUSY versions, such  as the 3-3-1 models, are  different because of the
     appearance of the $\mathrm{SU(3)_L}$ scale apart from the SUSY scale. For the SUSYE331, the parameters characterizing these two scales may  cancel each other to create light mass of the lightest singly charged Higgs  \cite{susye331r}.  In this work, we will investigate the DCHs
        in the SUSYRM331  and prove that there may exist  a light DCH, even if both soft
         and $\mathrm{SU(3)_L}$ parameters  are very large. Apart from inheriting the lepton number two,
   this light DCH is also lighter than  almost of  new particles in the model, therefore will decay mainly to same-sign lepton pair. So  the possibility of detection the lightest DCH will  increase at colliders such as  the LHC,  ILC and CLIC.
        In the left-right symmetric model, the cross
        sections for the DCH creation at the LHC are predicted below  5 fb for mass values
         being more than 200 GeV.
        In the SUSY left-right model they are estimated below 10 fb with collision energy is 14 TeV at LHC \cite{h2lr}
       and the mass of DCH is smaller than 450 GeV. The cross sections for the DCH creation  will decrease if their masses increase.
       In the framework of
        the 3-3-1 model, the cross sections for creating DCHs  can reach  the value smaller than  $10^{2}$ fb
        in $e^+e^-$ colliders \cite{h2331m,h2331m2}.  Our work will concentrate on the signals of detection DCHs
         at  the ILC and CLIC because  of their very high precision.
         In addition the collision energies of the ILC and CLIC are smaller than that of the LHC  but the   total cross
         sections for creating the DCHs at the ILC and CLIC are larger than those at the LHC.

Let us remind the reason for studying the
 3-3-1 models. The 3-3-1 models not only contain
the great success  of the SM but also can solve many  problems of the SM.
In particularly the 3-3-1 models can provide the neutrino small masses as well as candidates for the DM \cite{DM1,DM2}.
 The decays of some new particles
can solve the matter-antimatter asymmetry via leptogenesis
mechanisms \cite{lepto1,lepto2,lepto3}. The 3-3-1 models can connect to the cosmological inflation  \cite{lepto1,lepto2,lepto3}.  In addition
the 3-3-1 models
 \cite{pisano1,rfoot,331r1,331r2,331r3,331r4,331r5,331r6,orminimal} have many intriguing properties.
In order to make the models anomaly free, one of the  quark families must
transform under $\mbox{SU}(3)_L$ differently from  other two. This
leads to a consequence that the number of fermion generations has to
be multiple of the color number which is three. In combination
with the QCD asymptotic freedom requiring the number of quark
generations must be less than five, the solution is exactly three
for the number of fermion generations as required. Furthermore, the
3-3-1 models give a  good explanation of the electric charge
quantization \cite{ecq1,ecq2,ecq3,ecq4,ecq5}.

It is to be noted that the unique disadvantage of the 3-3-1 models
is the complication in the Higgs sector, which reduces their
predictability. Recently, there have been some efforts to
reduce the Higgs contents of the models. The first successful
attempt was to the 3-3-1 model with right-handed neutrinos
\cite{331r1,331r2,331r3,331r4,331r5,331r6} giving the model with just two Higgs triplets. The model is
called by the economical 3-3-1 model \cite{e3311,e3312,e3313}. The similar
version  to the minimal 3-3-1 model with Higgs sector containing
three triplets and one sextet is the reduced minimal 3-3-1 model
 with again just two Higgs triplets \cite{ffv1,ffv2}. However, to
give  masses to all fermions  in the 3-3-1 models with the minimal
Higgs sector, ones have to introduce the effective couplings which
are non-renormalizable.  On the other hand, 	by investigating the one-loop $\beta$-function in  the minimal 3-3-1 model
	and its supersymmetric 	version predict the existence of Landau poles that make these theories lose
	the perturbative character.  In order to solve this problem, the cut-off
  $\Lambda \simeq \mathcal{O}(1)$  TeV should be implied
  \cite{thema1,thema2}.  For  the non-suppersymmetric version, the upper bound of
   $\Lambda< 5$ TeV seems inconsistent with recent data of precision tests
   \cite{buras2,331d}. As a solution to this problem,  the  SUSY version predicts
   a less restrict  upper bound.  And the $\rho$ parameter, one of
the most important parameters for checking the precision test at
low energy \cite{stup1,stup2}, still satisfies the current data
if SUSY contributions are considered \cite{rhomssm}.  Anyway,  discussions for non-SUSY version predicts that the valid scale of the
SUSYRM331 should be large, resulting  the very heavy masses of the new particles, except
a light neutral CP-even Higgs and maybe the lightest DCH. Therefore, apart from the light neutral Higgs which can be identified with the one observed at LHC recently, the lightest DCH is the only one
 which may be observed by recent experiments.

 Once again we would like to emphasize that the RM331 model contains
  the minimal number of  Higgses, the first way to generate consistent masses
  for fermions is introducing  the effective  operators working at the TeV scale \cite{them1,Ferr}.
 Besides that,  in the  SUSY versions  the fermion masses  can be generated  by
 including the radiative corrections through the mixing of fermions and their superpartners \cite{them21,them22,huong1}.
  Of course, in this case the well-known $R$ parity has to be broken.
  Based on these results, many supersymmetric versions have been built and studied such as
SUSYE331  \cite{se331,se3311,se3312,se3313}, SUSYRM331 \cite{huong1,Ferr},... One of the intriguing
 features of supersymmetric theories is that the Higgs spectrum  is quite constrained.

Our paper is organized  as follows.  In  section
\ref{sec:sm}, we will briefly  review the SUSYRM331 model, specially
 concentrate on the Higgs, gauge boson sectors and
 their effect to the $\rho$ parameter, which may indirectly affect to the lower bound
 of the $SU(3)_L$ scale. Furthermore, some important and interesting properties of the SUSYRM331 are discussed,
 for  example: (i)   the soft and the $\mathrm{SU(3)_L}$ parameters
 should be in the same order; (ii)   the model contains a light neutral CP-even Higgs with the values of the
 squared tree-level mass of $m^2_Z|\cos 2\gamma|+ m^2_W\times \mathcal{O}(\epsilon)$.
  Here $\gamma$ is defined  as ratio of two vacuum expectation of two Higgses $\rho$
  and $\rho'$. While $\epsilon $ is defined as a quantity
 characterizing  the ratio of  the electroweak and $\mathrm{SU(3)_L}$ scales.
  Section  \ref{secdch} is devoted
  for investigation in details  the masses and  other properties of  the DCHs.
 We will discuss the constraint of the DCH masses under the recent experimental value of the decay of the lightest
  CP-even neutral Higgs to two photons.  From this we prove
  that there  exists a region of  parameter space containing a  light DCH.
    In Section \ref{signaldch} we discuss on the  creation of DCHs in
    the   $e^+e^-$  colliders such as the ILC
   and CLIC.  Specifically,   we  establish formulas of the cross sections for
   reactions  $e^+e^-\rightarrow H^{++}H^{--}$
   in collision energies from 1 to 3 TeV and calculate the number of events for
   the DCH creation.
    These cross sections  and
  the  Higgs masses are represented as functions of very convenient parameters such as  masses of
   neutral CP-odd Higgses,   mass of the heavy singly charged gauge boson,
   $\tan\gamma$ and $\tan\beta$  as ratios of Higgs vacuum expectation values (VEV),which will be defined  in the work.
   This will help one more easily  predict  many properties relating  to
   the DCHs as well as relations among  the masses of particles in the model. With each collision
   energy level of 1.5, 2 and 3 TeV, we discuss on
   the  parameter space where the masses of  three DCHs can satisfy the
   allowed kinetic condition, namely the  mass  of
   each DCH must be smaller than half of the collision energy.  Then we estimate the amplitudes
    of  the cross sections
   in these regions of parameter space. Finally,  the branching ratios of the  DCHs  decay to pairs of the
    same-sign leptons are briefly discussed.
\section{\label{sec:sm} Review of the SUSYRM331 model}

This  work bases on the models represented in \cite{huong1,Ferr}. For
convenience, we summarize important results which will be used in our calculation. Through the
work we will use the notation of
two-component spinor for fermions, where $\psi$ denotes for a particle and $\psi^c$ denotes
the corresponding anti-particle.
 Both $\psi$ and $\psi^c$ are left-handed  spinors. In case of the Majorana fields, where $\psi=\psi^c$,
 we will use  $\psi$ notation.

\subsection{Lepton and quark sectors}

 The  lepton sector is arranged based on the original non-supersymmetric version \cite{orminimal},
 namely
\begin{eqnarray}
 \hat{  L}_{l} =  \left(%
\begin{array}{ccc}
  \hat{ \nu}, &  \hat{ l }, & \hat{ l}^{c} \\
\end{array}%
\right)^T\sim ({\bf1},{\bf3},0), \,\ l= e, \mu , \tau.
\label{trip}
\end{eqnarray}
In parentheses it appears the transformation properties under the
respective factors $(\mathrm{SU(3)}_C,$
$~\mathrm{SU(3)}_L,~\mathrm{U(1)}_X)$.

In the quark sector, the first  quark family   is  put in a superfield
which transforms as a triplet  of  the $ \mathrm{SU(3)}_L$  group,
\bea
  \hat{ Q}_{1L}  =\left(%
\begin{array}{ccc}
  \hat {u}_{1}, & \hat{ d}_{1},& \hat{J_1 } \\
\end{array}%
\right)\sim \left({\bf3},{\bf3},\frac{2}{3}\right). \label{q1l}
\eea
Three respective anti-quark superfields are singlets of the $ \mathrm{SU(3)}_L$  group,
\be
\hat{u}^{c}_{1} \sim
\left({\bf3}^*,{\bf1},-\frac{2}{3}\right),\quad \hat{d}^{c}_{1}
\sim \left({\bf3}^*,{\bf1},\frac{1}{3}\right),\quad
\hat{J}^{c}_{1} \sim \left({\bf3}^*,{\bf1},-\frac{5}{3}\right).
\label{q1r}
\ee
The  remaining two quark families are included in two corresponding superfields  transforming  as
 antitriplets,
\be
\hat{ Q}_{iL} = \left(%
\begin{array}{ccc}
 \hat{d}_{i}, & - \hat{ u}_{i} & \hat{ j}_{i} \\
\end{array}%
\right)^T \sim \left({\bf3},{\bf3}^{*},-\frac{1}{3}\right), \,
i = 2, 3.  \label{q23l}
\ee
and  the respective anti-quark superfields are singlets,
\bea
\hat{u}^{c}_{i} \sim
\left({\bf3}^*,{\bf1},-\frac{2}{3}\right),\quad \hat{d}^{c}_{i}
\sim \left({\bf3}^*,{\bf1},\frac{1}{3}\right), \quad
\hat{j}^{c}_{i} \sim \left({\bf3}^*,{\bf1},\frac{4}{3} \right), i = 2, 3.  \label{q23r}
\eea
The SUSYRM331 needs four  Higgs superfileds in order to generate all masses of leptons and quarks, but  radiative corrections \cite{huong1} or effective operators \cite{Ferr} must be added.  For convenience in investigating the couplings between leptons and DCHs, in this work we will use the effective approach.
\subsection{\label{apllv} Gauge bosons and lepton-lepton-gauge boson vertices}
 The gauge boson sector of the SUSYRM331 model was thoroughly investigated in \cite{huong1,Ferr} and
this sector is similar to that of the non-SUSY version \cite{ffv1}. According to these works, the gauge
sector includes three neutral gauge bosons ($A,~Z, ~Z'$), four singly charged ($W^{\pm},~V^{\pm}$)
 and two doubly charged $U^{\pm \pm}$ gauge bosons.  Among them, $A, ~Z$ and $W^{\pm}$ are
  SM particles  while the remain  are $\mathrm{SU(3)}_L$ particles with masses being at the $\mathrm{SU(3)}_L$ scale.
 The new charged gauge bosons $V^{\pm}$ and $U^{\pm \pm}$ carry the  lepton numbers two, hence they are also
called by bilepton.  According analysis in \cite{dng}, mass of the
charged bilepton $U$  is always less than $0.5  m_{Z'}$.
Therefore, we expect that the decays $ Z' \longrightarrow  U^{++}
U^{--} $  and $U^{\pm \pm} \longrightarrow 2 l^{\pm} (l=e, \mu,
\tau) $ are allowed, leading to spectacular signals in the
future colliders. The DCHs also are bilepton, leading to a very interesting consequence:
the lightest DCH may be the lightest bilepton, it only decays to a charged lepton pair.
This is exactly the case happening in the SUSYRM331, as we will prove through this work. All  masses of the of gauge bosons can be written as
functions of $W$ and  $V$ gauge boson masses. There is a  simple relation   between $m_W,~m_{V}$ and $m_{U}$,
namely  $m^2_{U}=m^2_{W}+m^2_{V}$ which will be summarized in the Higgs sector.   Therefore, we can define $m_V$ as a parameter characterized for
the $\mathrm{SU(3)}_L$ scale. Recently,  the studies of flavor neutral changing current processes  and the muon anomalous
magnetic moment in the Reduced Minimal 3-3-1 Model
\cite{Kelso,cogo} have set the below limits of $m_V$,  namely
$m_V\geq 650$ and 910 GeV, respectively.

  The vertex of $ffV$, which is very important to studying the creation of DCHs in the $e^+e^-$ colliders, is represented  in the Lagrangian shown in
\cite{huong1,Ferr,ffv1,ffv2}, namely
\bea \mathcal{L}_{ffV}=
g\bar{L}\bar{\sigma}^{\mu}\frac{\lambda^a}{2}
 L V^a_{\mu}. \eea  The relations between mass  and original states
of neutral gauge bosons are given as follows
\bea \left(%
\begin{array}{c}
  W_{3\mu} \\
  W_{8\mu} \\
  B_{\mu} \\
 \end{array}%
\right)=\mathcal{C}_\mathrm{B} \left(%
\begin{array}{c}
  A_{\mu} \\
  Z_{\mu} \\
  Z'_{\mu} \\
 \end{array}%
\right)= \left(%
\begin{array}{ccc}
  \frac{t}{\sqrt{2(2t^2+3)}}, & \frac{\sqrt{3}}{2}\left(c_{\zeta}+\frac{s_{\zeta}}{\sqrt{2t^2+3}}\right),
   & \frac{\sqrt{3}}{2}\left(-s_{\zeta}+\frac{c_{\zeta}}{\sqrt{2t^2+3}}\right) \\
   -\frac{\sqrt{3}t}{\sqrt{2(2t^2+3)}}& \frac{1}{2}\left(c_{\zeta}-\frac{3s_{\zeta}}{\sqrt{2t^2+3}}\right),
   &  -\frac{1}{2}\left(s_{\zeta}+\frac{3c_{\zeta}}{\sqrt{2t^2+3}}\right)\\
    \frac{\sqrt{3}}{\sqrt{2t^2+3}}&-\frac{\sqrt{2}s_{\zeta}t}{\sqrt{2t^2+3}} &-\frac{\sqrt{2}c_{\zeta}t}{\sqrt{2t^2+3}}\\
\end{array}%
\right) \left(%
\begin{array}{c}
  A_{\mu} \\
  Z_{\mu} \\
  Z'_{\mu} \\
 \end{array}%
\right).\crn\label{bosonstate1}\eea
 Here $c_{\zeta}\equiv\cos\zeta>0,~ s_{\zeta}\equiv\sin\zeta>0$ with $\zeta$  satisfying
 \be  \tan2\zeta\equiv\frac{\sqrt{(3 + 2 t^2) }(m_V^2 - m_W^2)}{(1+t^2) (m^2_V
 +m_W^2)}.\label{dzeta}\ee
 The parameter $t$ is the ratio between $g'$ and $g$, namely
 \be t\equiv \frac{g'}{g}=\sqrt{\frac{6\sin^2\theta_W}{1-4\sin^2\theta_W}}.\label{tvalues}  \ee
Masses of gauge bosons are  given by
\bea m_{\gamma}&=&0,\crn m^2_{Z}&=& \frac{t^2+2}{3}\left(
m^2_U-\sqrt{m_U^4-\frac{4(2t^2+3)}{(t^2+2)^2}m^2_Vm^2_W}\right),\crn
 m^2_{Z'}&=& \frac{t^2+2}{3}\left(
m^2_U+\sqrt{m_U^4-\frac{4(2t^2+3)}{(t^2+2)^2}m^2_Vm^2_W}\right).
\label{bosonmass1}\eea The $Z-Z'$ mixing angle in the
framework of the RM331 model is quite small $|\phi | <
10^{-3}$ \cite{ffv2}. It is interesting to note that, due to the generation
discrimination in the 3-3-1 models,   the new neutral gauge boson
$Z'$ has the flavor changing neutral current \cite{fcnc1,fcnc2,fcnc3}.

Above analysis is enough to calculate vertex factors of charged
  leptons with neutral gauge bosons, as shown explicitly in Table
  \ref{eeVvertices1}. Here we only concentrate on the largest vertex couplings by assuming that the flavor
  basis of leptons and quarks  is the mass basis.
\begin{table}[h]
  \centering
  \begin{tabular}{|c|c|c|c|}
    \hline
    $\bar{f}fV_{\mu}$ &  $A_{\mu}$ &$Z_{\mu}$ &$Z'_{\mu}$ \\
     \hline
     $\nu_{e}$ & $0$  & $\frac{igc_{\zeta}}{\sqrt{3}}\bar{\sigma}^{\mu}$ &
      $\frac{-igs_{\zeta}}{\sqrt{3}}\bar{\sigma}^{\mu}$\\
        \hline
    $e$&  $-ie$ & $-\frac{ig}{2\sqrt{3}}\left( c_{\zeta}+\frac{3s_{\zeta}}{\sqrt{2t^2+3}}\right)
    \bar{\sigma}^{\mu}$ & $\frac{ig}{2\sqrt{3}}\left( s_{\zeta}-\frac{3c_{\zeta}}{\sqrt{2t^2+3}}\right)
    \bar{\sigma}^{\mu}$ \\
    \hline
     $e^c$ & $ie$  & $-\frac{ig}{2\sqrt{3}}\left( c_{\zeta}-\frac{3s_{\zeta}}{\sqrt{2t^2+3}}\right)
    \bar{\sigma}^{\mu}$  &
      $\frac{ig}{2\sqrt{3}}\left( s_{\zeta}+\frac{3c_{\zeta}}{\sqrt{2t^2+3}}\right)
    \bar{\sigma}^{\mu}$ \\
     \hline
     $u$ & $\frac{i2e}{3}$  & $\frac{ig}{\sqrt{3}}\left( c_{\zeta}-\frac{2t^2s_{\zeta}}{3\sqrt{2t^2+3}}\right)
    \bar{\sigma}^{\mu}$  &
      $\frac{-ig}{\sqrt{3}}\left( s_{\zeta}+\frac{2t^2c_{\zeta}}{3\sqrt{2t^2+3}}\right)
    \bar{\sigma}^{\mu}$ \\
    \hline
     $u^c$ & $-\frac{i2e}{3}$  &
     $\frac{2igt^2s_{\zeta}}{3\sqrt{3(2t^2+3)}}
    \bar{\sigma}^{\mu}$  &
       $\frac{2igt^2 c_{\zeta}}{3\sqrt{3(2t^2+3)}}
    \bar{\sigma}^{\mu}$ \\
     \hline
     $d$ & $-\frac{ie}{3}$  & $-\frac{ig}{2\sqrt{3}}\left( c_{\zeta}+\frac{(4t^2+9)s_{\zeta}}{3\sqrt{2t^2+3}}\right)
    \bar{\sigma}^{\mu}$  &
      $\frac{ig}{2\sqrt{3}}\left( s_{\zeta}-\frac{(4t^2+9)c_{\zeta}}{3\sqrt{2t^2+3}}\right)
    \bar{\sigma}^{\mu}$\\
    \hline
     $d^c$ & $\frac{ie}{3}$  &
     $-\frac{igt^2s_{\zeta}}{3\sqrt{3(2t^2+3)}}
    \bar{\sigma}^{\mu}$  &
       $-\frac{igt^2 c_{\zeta}}{3\sqrt{3(2t^2+3)}}
    \bar{\sigma}^{\mu}$ \\
     \hline
     $J_1$ & $\frac{5ie}{3}$  & $-\frac{ig}{2\sqrt{3}}\left( c_{\zeta}+\frac{(4t^2-9)s_{\zeta}}{3\sqrt{2t^2+3}}\right)
    \bar{\sigma}^{\mu}$  &
      $\frac{ig}{2\sqrt{3}}\left( s_{\zeta}+\frac{(4t^2-9)c_{\zeta}}{3\sqrt{2t^2+3}}\right)
    \bar{\sigma}^{\mu}$\\
    \hline
     $J_1^c$ & $-\frac{5ie}{3}$  &
     $\frac{5igt^2s_{\zeta}}{3\sqrt{3(2t^2+3)}}
    \bar{\sigma}^{\mu}$  &
       $\frac{5igt^2 c_{\zeta}}{3\sqrt{3(2t^2+3)}}
    \bar{\sigma}^{\mu}$ \\
    \hline
     $c,~t$ & $\frac{i2e}{3}$  & $\frac{ig}{2\sqrt{3}}\left( c_{\zeta}+\frac{(2t^2+9)s_{\zeta}}{3\sqrt{2t^2+3}}\right)
    \bar{\sigma}^{\mu}$  &
      $\frac{-ig}{2 \sqrt{3}}\left( s_{\zeta}-\frac{(2t^2+9)c_{\zeta}}{3\sqrt{2t^2+3}}\right)
    \bar{\sigma}^{\mu}$ \\
    \hline
     $c^c,~t^c$ & $-\frac{i2e}{3}$  &
     $\frac{2igt^2s_{\zeta}}{3\sqrt{3(2t^2+3)}}
    \bar{\sigma}^{\mu}$  &
       $\frac{2igt^2 c_{\zeta}}{3\sqrt{3(2t^2+3)}}
    \bar{\sigma}^{\mu}$ \\
    \hline
     $s,~b$ & $-\frac{ie}{3}$  & $-\frac{ig}{\sqrt{3}}\left( c_{\zeta}+\frac{t^2 s_{\zeta}}{3\sqrt{2t^2+3}}\right)
    \bar{\sigma}^{\mu}$  &
      $\frac{ig}{\sqrt{3}}\left( s_{\zeta}-\frac{4t^2c_{\zeta}}{3\sqrt{2t^2+3}}\right)
    \bar{\sigma}^{\mu}$\\
    \hline
     $s^c,~b^c$ & $\frac{ie}{3}$  &
     $-\frac{igt^2s_{\zeta}}{3\sqrt{3(2t^2+3)}}
    \bar{\sigma}^{\mu}$  &
       $-\frac{igt^2 c_{\zeta}}{3\sqrt{3(2t^2+3)}}
    \bar{\sigma}^{\mu}$ \\
    \hline
     $j_1,~j_2$ & $-\frac{4ie}{3}$  & $-\frac{ig}{2 \sqrt{3}}\left( c_{\zeta}+\frac{(2t^2-9) s_{\zeta}}{3\sqrt{2t^2+3}}\right)
    \bar{\sigma}^{\mu}$  &
      $-\frac{ig}{2\sqrt{3}}\left( s_{\zeta}-\frac{(2t^2-9)c_{\zeta}}{3\sqrt{2t^2+3}}\right)
    \bar{\sigma}^{\mu}$\\
    \hline
     $j_1^c,~j^c_2$ & $\frac{4ie}{3}$  &
     $-\frac{4igt^2s_{\zeta}}{3\sqrt{3(2t^2+3)}}
    \bar{\sigma}^{\mu}$  &
       $-\frac{4igt^2 c_{\zeta}}{3\sqrt{3(2t^2+3)}}
    \bar{\sigma}^{\mu}$ \\
    \hline
  \end{tabular}
  \caption{Vertex factors between leptons, quarks and neutral gauge bosons.
  Note that $e=g\sin\theta_W$.}\label{eeVvertices1}
\end{table}

 \subsection{Constraint from $\rho$ parameter}
The above analysis shows that  the  structure of the neutral gauge bosons is
the same as that of the RM331 model when all mixing and mass parameters of
these bosons are written in term of the charged gauge boson masses. So the
contributions of new heavy gauge bosons to the $\rho$ parameter from
the $SU(3)_L$ charged gauge bosons are given in \cite{rhom1,rhom2}.
They also relate to the $T$ parameter through the equality $\Delta\rho\equiv \rho-1
\simeq \widehat{\alpha}(m_Z) T$, where $\widehat{\alpha}(m_Z)$ is the fine structure
 constant defined in the minimal scheme ($\overline{MS}$) at the $m_Z$ scale
\cite{pdg2014}. The problem is that all of these contributions are always positive
so the total always increase the value of $\Delta\rho$ larger than the current
experimental upper bound, unless the $SU(3)_L$ scale is larger than 9 TeV \cite{331d}.

Because the new quarks are $SU(2)_L$ singlets, they do not
contribute to the $\rho$ parameters. The other contributions arise
 from Higgses and SUSY particles including  Higgsinos, gauginos and
 superpartners of the fermions. Being the functions of SUSY parameters
 they  are completely independent on the $SU(3)_L$ scale.
The contributions of the DCHs are from only couplings \cite{rhomssm}
\be i c \phi^*_1\partial_{\mu}\phi_2 V^{\mu}+h.c \hs (V=W,Z)\label{cHHv}
\ee of two charged Higgses $\phi_1$ and $\phi_2$.
According to the table \ref{hcc3coupling},
there are only non-zero vertices of
$\chi^{++}W^-H^{-}_2$ and $\chi^{\prime++}W^-H^{-}_2$
 relating with DCHs.  Because $\chi^{\pm\pm}$ and $\chi'^{\pm\pm}$
 contribute to mainly to $H^{\pm\pm}_1$ and the goldstone of $U^{\pm}$ boson,
 they mix with other two DCHs with  very small factors of orders being smaller
 than $\mathcal{O}\left(\frac{m_W^3}{m_V^3}\right)$. In addition, the kind of the interactions given in
  (\ref{cHHv}) with two identical DCHs  gives zero contribution to the $\rho$ parameter \cite{rhomssm}.
 So the total contribution of the physical DCHs to the $\rho$ parameter is very
 suppressed.

\begin{figure}[h]
  \centering
 \includegraphics[width=10cm]{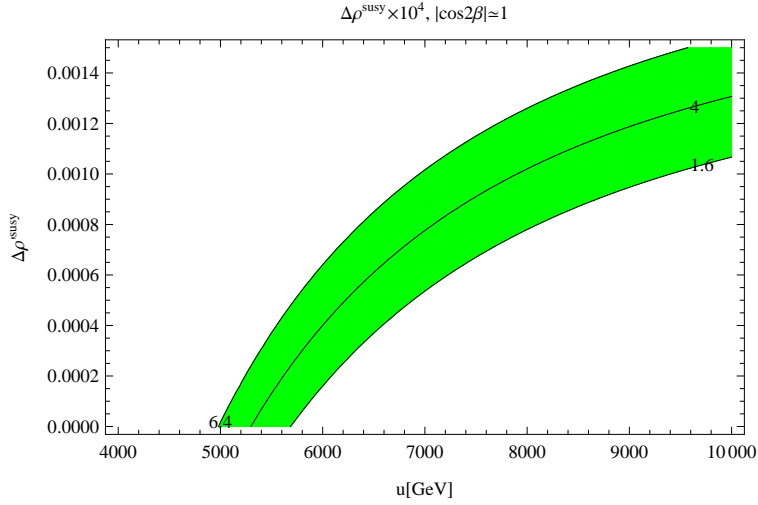}
    \caption{ Contour plot of $\Delta\rho^{\mathrm{susy}}$ as function of $\Delta\rho^{\prime\mathrm{susy}}$ and   $SU(3)_L$ scale $u$. The green region  satisfies $1.6\times 10^{-4}\leq\Delta\rho^{\mathrm{susy}}\leq 6.4\times 10^{-4}$. }\label{rhoSUSYRM331}
\end{figure}

 Because  the  Higgs triplets $\rho$ and $\rho'$ break $SU(2)_L$ symmetry, they will give
 main contributions  to the couplings of singly and neutral Higgses to  normal
 gauge bosons and therefore may affect significantly  to the $\rho$ parameter. This is very
 similar to the case of the MSSM. In fact, the SUSYRM331 contains
 two CP-even neutral Higgses and two  singly charged Higgses $H^{\pm}_1$
 behaving the same as those in the MSSM.  More explicitly, they couple with
 $W$ and $Z$ boson the in same way as those in the MSSM, especially in the  large limit
of the $SU(3)_L$ and soft SUSY breaking scales, which is
 exactly the valid condition of the SUSYRM331. So the  total contribution
 to the $\rho$ parameter of these SUSYRM331 Higgses is nearly the same as what
 found in the MSSM. In general, the  contributions to the $\rho$ parameter
 obtained from the investigation for  MSSM can also be used for the SUSY331
 version  \cite{rhomssm}. The most important results are: i) all unexpected positive
 contributions decrease rapidly to zero when the overall sparticle mass scale is
 large enough,  ii) the negative contribution from the Higgs scalars can reach the
 absolute values of $10^{-4}$, which is the order of  the recent sensitive experimental
 value of the $\rho$ parameter. In the SUSYRM331 framework,  the total positive
 SUSY contribution can be set to the order of $\mathcal{O}(10^{-4})$, because
 the soft parameters are at least in the order of the $SU(3)_L$ scale, i.e the
 TeV scale. While the total contribution from Higgs scalar is completely different.
 It has negative value when the masses of CP-odd neutral Higgses are very large and
 the lightest CP-even neutral Higgs reaches  the largest value of $M_{Z}|\cos2\beta|$
 at the tree level \footnote{Note that the well-known $\beta$ in the MSSM is different
 from our definition of $\beta$ in (\ref{dn1}). In fact, the $\gamma$ parameter in
  (\ref{dn1}) plays the same role as the $\beta$ in the MSSM}\cite{rhomssm,adjo}.
  Then, the contribution from the Higgs sector is
\bea  \Delta\rho^{\mathrm{susy}}_{\mathrm{H}}= \frac{3\alpha}{16\pi^2\sin^2\theta_W}
f_{H}(\cos^22\beta,\theta_W),  \label{drhohiggs} \eea
where $$ f_H(x,\theta_W)\equiv x\left( \frac{\ln\left(\cos^2\theta_W/x\right)}{\cos^2\theta_W-x}+  \frac{\ln x}{\cos^2\theta_W(1-x)}\right).$$
In the following we will show that the negative
$\Delta\rho^{\mathrm{susy}}_{\mathrm{H}}$ can cancel the new positive contributions
arising from the SUSY and 3-3-1 properties. The total deviation of the $\rho$ parameter
can be divided into three parts,
\be \Delta\rho^{\mathrm{susy}}= \widehat{\alpha}(m_Z) T_{\mathrm{min}}+  \Delta\rho^{\mathrm{susy}}_{\mathrm{H}}+ \Delta\rho^{\prime\mathrm{susy}},\label{delrho1} \ee
where $\Delta\rho^{\prime\mathrm{susy}}$ is the total positive contribution of the Higgsino, gaugino and sfermion particles;  $T_{\mathrm{min}}$ is the  contribution from the minimal 331 framework to the oblique $T$ parameter\cite{rhom1,rhom2},
\bea  T_{\mathrm{min}} &=& \frac{3\sqrt{2}G_F}{16\pi^2 \widehat{\alpha}(m_Z)}\left[ m^2_U+m^2_V-\frac{2m^2_Um^2_V}{m^2_U-m^2_V}\ln\frac{m^2_U}{m^2_V}\right]\crn
&+& \frac{1}{4\pi \sin^2\theta_W}\left[ 2-\frac{m^2_Um^2_V}{m^2_U-m^2_V}\ln\frac{m^2_U}{m^2_V}+3\tan^2\theta_W\ln\frac{m^2_U}{m^2_V} \right]+\frac{m^2_{Z}-m_{Z_0}^2}{\widehat{\alpha}(m_Z)m^2_{Z}}, \label{tmin}\eea
where $m_{Z}, m_{Z'}, m_U$ and $m_V$ are the masses of gauge bosons predicted by the SUSYRM331.  All of the experimental values are given in  \cite{pdg2014}, namely $m_{Z_0}=91.1876\pm0.0021$ GeV, $m_W=80.385\pm 0.015$ GeV, $\sin^2\theta_W=0.23126$, $G_F=1.166 3878(6)\times 10^{-5}\mathrm{GeV^{-2}}$, and $ \widehat{\alpha}^{-1}(m_Z)=127.940\pm0.014$.  Also, the experimental constraint of new physics to $\Delta\rho$ is $1.6\times 10^{-4}\leq\Delta\rho\leq 6.4\times 10^{-4}$ \cite{pdg2014}.
The $\Delta\rho^{\mathrm{susy}}$ now is the function of $|\cos2\beta|$, $\Delta\rho^{\prime\mathrm{susy}}$ and the $SU(3)_L$ scale
$u=\sqrt{w^2+w'^2}$. With the discovery of the neutral CP-even Higgs with mass of 125 GeV, the $\beta$ should satisfies $|\cos{2\beta}|\rightarrow 1$. The numerical result of $\Delta\rho^{\mathrm{susy}}$ is shown
in the figure \ref{rhoSUSYRM331}, where the lower bound of $u \geq 5$ TeV is allowed.

Finally, what we stress here is that the sum of the respective negative and positive contributions from $\Delta\rho^{\mathrm{susy}}_{H}$ and $\Delta\rho^{\prime\mathrm{susy}}$ is
enough to keep the value of the $\rho$ parameter in the allowed constraint. Therefore,  being different
from the non-SUSY version, in the SUSY view  the $SU(3)_L$ scale is free from the
constraint of the $\rho$ parameter.

 On the other hand, the $SU(3)_L$ scale is constrained by investigating the $Z'$ boson. According to (\ref{bosonmass1}) we get
 $$m_{Z'}\simeq\frac{2m_V c_W }{ \sqrt{3(1-4\sin^2\theta_W)}}$$ in the limit of $u\gg v,v'$.

In the frame work of the minimal 3-3-1 models, the investigation of  the  LEP-II constraints on $m_{Z'}$ \cite{buras2} as well as the  $B_d\rightarrow K^* \mu\mu$
data at LHC indicated that the   lower bounds of $m_{Z'}$ must be above  7 TeV \cite{zp0,zp1,zp2}. In addition, the above discussion suggests that the $Z'$ boson in the SUSYRM331 model behaves similarly to the one in the non-SUSY version at the tree level. Combining with the constraint of $m_{Z'}$ in order to avoid  the Landau pole,  the SUSYRM331 model predicts the most interesting range of $m_{Z'}$ is $7\;\mathrm{TeV} \leq m_{Z'}\leq 9$ TeV, leading to $2\;\mathrm{TeV}\leq m_V\leq 3\; \mathrm{TeV}$.

\subsection{Higgs sector}

The scalar superfields, which are necessary to
generate the fermion masses, are
\begin{equation}
\hat{\rho} =
      \left( \begin{array}{c} \hat{\rho}^{+} \\
                 \hat{ \rho}^{0} \\
                  \hat{\rho}^{++}          \end{array} \right) \sim
({\bf1},{\bf3},+1),\quad \hat{\chi} =
      \left( \begin{array}{c} \hat{ \chi}^{-} \\
                \hat{  \chi}^{--} \\
                \hat{  \chi}^{0}          \end{array} \right) \sim
({\bf1},{\bf3},-1). \label{3t}
\end{equation}
To remove chiral anomalies generated by the superpartners of the
scalars,   two new scalar superfields are introduced to transform as anti-triplets
under the $SU(3)_L$, namely
\begin{equation}
\hat{\rho}^{\prime} =
      \left( \begin{array}{c} \hat{\rho}^{\prime-} \\
                  \hat{\rho}^{\prime0} \\
                  \hat{\rho}^{\prime--}          \end{array} \right)
\sim ({\bf1},{\bf3}^{*},-1),\quad \hat{\chi}^{\prime} =
      \left( \begin{array}{c} \hat{\chi}^{\prime+} \\
                  \hat{\chi}^{\prime++} \\
                  \hat{\chi}^{\prime0}          \end{array} \right)
\sim ({\bf1},{\bf3}^{*},+1). \label{shtc}
\end{equation}
The pattern of the symmetry breaking of the model is given by the
following scheme (using the notation given at \cite{massspectrum})
\bea \mbox{SUSY RM
3-3-1}&\stackrel{\mathcal{L}_{\mathrm{soft}}}{\longmapsto}&
\mbox{SU(3)}_C\ \otimes \ \mbox{SU(3)}_{L}\otimes
\mbox{U(1)}_{X}\crn&\stackrel{\langle\chi\rangle \langle
\chi^{\prime}\rangle}{\longmapsto}& \mbox{SU(3)}_{C} \ \otimes \
\mbox{SU(2)}_{L}\otimes \mbox{U(1)}_{Y} \crn
&\stackrel{\langle\rho\rangle \langle
\rho^{\prime}\rangle}{\longmapsto}& \mbox{SU(3)}_{C} \ \otimes \
\mbox{U(1)}_{Q}. \label{breaksusy331tou1} \eea For the sake of
simplicity, all vacuum expectation values (VEVs)
are supposed to be real.  When the 3-3-1 symmetry is broken, i.e,
$\mathrm{SU(3)}_{C} \otimes \mathrm{U(1)}_{Q}$,   VEVs of the  scalar
 fields are defined as follows
\begin{eqnarray}
\langle \rho \rangle &=& \left(%
\begin{array}{ccc}
  0, & \frac{v}{\sqrt{2}}, & 0 \\
\end{array}%
\right)^T,\quad \langle \chi \rangle = \left(%
\begin{array}{ccc}
  0, & 0, & \frac{w}{\sqrt{2}} \\
\end{array}%
\right)^T, \crn \langle \rho^{\prime} \rangle &=&
      \left(%
\begin{array}{ccc}
  0, & \frac{v'}{\sqrt{2}}, & 0 \\
\end{array}%
\right)^T,\quad \langle \chi^{\prime} \rangle =
      \left(%
\begin{array}{ccc}
  0, & 0, & \frac{w'}{\sqrt{2}} \\
\end{array}%
\right)^T. \label{vev1}
\end{eqnarray}
 Because the symmetry breaking happens through steps given in (\ref{breaksusy331tou1}),
 the VEVs   have to satisfy the condition $ w,~w^{\prime} \gg v,~v{^\prime}$.
  The constraint on the $W$ bosons mass leads to the
  consequence that \begin{equation}
V^{2}\equiv v^{2}+v^{\prime 2}=(246\;{\rm GeV})^2.
\label{wmasslimite}
\end{equation}

\subsection{\label{Higgs}Higgs spectrum}

As usual, the scalar Higgs potential is written as in \cite{huong1}, except $V_\mathrm{soft}$
which is added $b$-type terms \cite{Ferr} to guarantee the vacuum stability of the model and
 to avoid the appearance of many tachyon scalars \cite{10,Barat1}.
 Therefore we have
 \be
V_{\mathrm{SUSYRM}331}=V_{\mathrm{D}}+V_{\mathrm{F}}+V_{\mathrm{soft}}
\label{ep1} \ee with
\bea
V_{\mathrm{D}}&=&-\mathcal{L}_{D}=\frac{1}{2}\left(D^{a}D^{a}+DD\right)\crn
&=& \frac{g^{\prime
2}}{12}(\bar{\rho}\rho-\bar{\rho^{\prime}}\rho^{\prime}
-\bar{\chi}\chi+\bar{\chi^{\prime}}\chi^{\prime})^{2} +
\frac{g^{2}}{8}\sum_{i,j}\left(\bar{\rho}_{i}\lambda^{a}_{ij}\rho_{j}
+\bar{\chi}_{i}\lambda^{a}_{ij}\chi_{j}-
\bar{\rho^{\prime}}_{i}\lambda^{*a}_{ij}\rho^{\prime}_{j}
-\bar{\chi^{\prime}}_{i}\lambda^{*a}_{ij}\chi^{\prime}_{j}
\right)^{2}, \crn
V_{\mathrm{F}}&=&-\mathcal{L}_{F}=\sum_{F}\bar{F}_{\mu}F_{\mu}
 = \sum_{i}\left[ \left\vert
\frac{\mu_{\rho}}{2}\rho^{\prime}_{i}\right\vert^{2}+ \left\vert
\frac{\mu_{\chi}}{2}\chi^{\prime}_{i}\right\vert^{2} + \left\vert
\frac{\mu_{\rho}}{2}\rho_{i} \right\vert^{2}+ \left\vert
\frac{\mu_{\chi}}{2}\chi_{i} \right\vert^{2} \right], \crn
V_{\mathrm{soft}}&=&-\mathcal{L}_{\mathrm{SMT}}=
m^{2}_{\rho}\bar{\rho}\rho+ m^{2}_{\chi}\bar{\chi}\chi+
m^{2}_{\rho^{\prime}}\bar{\rho^{\prime}}\rho^{\prime}+
m^{2}_{\chi^{\prime}}\bar{\chi^{\prime}}\chi^{\prime}-
\left(b_{\rho}\rho\rho'+b_{\chi}\chi\chi'+ \mathrm{h.c.}\right),
\label{ess} \eea
 where  $m_{\rho},~ m_{\chi},~ m_{\rho^{\prime}}$ and $m_{\chi^{\prime}}$ have the mass
 dimension.  Both $b_{\rho}$  and
$b_{\chi}$  have squared mass dimension and are assumed to be real  and positive to make sure the non-zero and real values of VEVs.
 The expansions of the neutral scalars
around their VEVs are
\begin{eqnarray}
\langle \rho \rangle&=&\frac{1}{\sqrt{2}}
      \left( \begin{array}{c} 0 \\
v+H_{\rho}+iF_{\rho} \\
                  0          \end{array} \right), \,
\langle \rho^{\prime}  \rangle=\frac{1}{\sqrt{2}}
      \left( \begin{array}{c} 0 \\
v'+H_{\rho^{\prime}}+iF_{\rho^{\prime}}\\
                  0          \end{array} \right), \, \crn
\langle \chi  \rangle&=&\frac{1}{\sqrt{2}}
      \left( \begin{array}{c} 0 \\
                  0 \\
w+H_{\chi}+iF_{\chi} \end{array} \right),\, \langle \chi^{\prime}
 \rangle=\frac{1}{\sqrt{2}}
      \left( \begin{array}{c} 0 \\
                  0 \\
w'+H_{\chi^{\prime}}+iF_{\chi^{\prime}}\end{array} \right).
\label{develop}
\end{eqnarray}
 The minimum of the Higgs potential corresponds to the vanishing  of all linear Higgs terms
  in the above potential.  As a result, it leads
 to four independent equations shown in \cite{Ferr} which reduce four independent parameters
  in the original  Higgs potential. We will use notations
 chosen in \cite{huong1} for this work. Especially, two  independent parameters are chosen as
 \be t_{\gamma}=\tan\gamma=\frac{v}{v'}, \hs t_{\beta}=\tan\beta=\frac{w}{w'}. \label{dn1}\ee
 They are two ratios of VEVs of neutral Higgs scalars, and similar to the $\beta$ parameter defined in the MSSM.
The two  electroweak and
 $\mathrm{SU(3)}_L$ scales relate with masses of $W$ and $V$ boson \cite{huong1,Ferr} by two equations
\begin{eqnarray}
  m_W^2 &=& \frac{g^2}{4}(v^2+v'^2); \hs
  m_V^2=\frac{g^2}{4}(w^2+w'^2).\
\nn \end{eqnarray}
 We can choose $m_V$ as   an independent parameter.  On the other hand, there are two heavy doubly charged
  bosons, denoted as $U^{\pm\pm}$, with mass $m_U$ satisfying
 $m^2_U=m^2_V+m^2_W$.  If $m_V\gg m_W$, there will appear a degeneration
  of two heavy boson masses,
 $m_U=  m_V+ \frac{1}{2} m_W \times \mathcal{O}\left(m_W/m_V\right)+
  m_W \times \mathcal{O}\left(m_W/m_V\right)^3 \simeq m_V$.
 As mentioned  above,  the  constraint of $m_V$   gives  a very small
   ratio between two scales $SU(2)_L$ and
    $SU(2)_L$:  $m^2_W/m^2_V\leq \mathcal{O}(10^{-3})$.  This is a rather good limit for
   our approximation used in  this  work.  The minimum conditions of the superpotential  result a series  of  the four below  equations
\begin{eqnarray}
  m_\rho^2+\frac{1}{4}\mu_\rho^2 &=& \frac{b_\rho}{t_\gamma}-\frac{1+t^2}{3}\times
  m_V^2\cos{2\beta} +\frac{t^2+2}
  {3}\times m_W^2\cos{2\gamma}, \label{linear2}\\
  m_\chi^2+\frac{\mu_\chi^2}{4} &=&
  \frac{b_\chi}{t_\beta}-\frac{2+t^2}{3}\times m_V^2\cos{2\beta}
  +\frac{1+t^2}{3}\times m_W^2\cos{2\gamma}, \label{linear3} \\
  s_{2\gamma}\equiv\sin{2\gamma}&=&\frac{2b_\rho}{m_\rho^2+m_{\rho'}^2+\frac{1}{2}\mu_\rho^2},\hs
  s_{2\beta}\equiv\sin{2\beta}=\frac{2b_\chi}{m_\chi^2+m_{\chi'}^2+\frac{1}{2}\mu_\chi^2} \label{linear4}
\end{eqnarray}
The two equations in (\ref{linear4}) show the relations between soft-parameters and ratios of the VEVs, and they are much the same as shown in the MSSM.  To estimate the scale of these soft parameters, based on the calculation in \cite{susye331r}
it is useful to write  two equations (\ref{linear2}) and (\ref{linear3}) in the new forms as follows
\be
 c_{2\gamma}\equiv\cos{2\gamma} = \frac{-( m_\chi^2+\frac{\mu_\chi^2}{4}-\frac{b_\chi}{t_\beta})(1+2s^2_W)
  +2(m_\rho^2+\frac{\mu^2_\rho}{4}-\frac{b_\rho}{t_\gamma})c^2_{W}}{m^2_W},\label{c2gamma}
\ee
\begin{eqnarray}
 c_{2\beta}\equiv\cos{2\beta}
&=&\frac{(m_\rho^2+\frac{\mu^2_\rho}{4}-\frac{b_\rho}{t_\gamma})(1+2s^2_W)-
2(m_\chi^2+\frac{\mu_\chi^2}{4}-\frac{b_\chi}{t_\beta})c^2_W}{m^2_V}\crn
&=& \frac{m_W^2}{m_V^2}\times \frac{(1+3t^2_W)c_{2\gamma}}{2} - \frac{3(1-4 s^2_W)}{2c^2_W}\times \frac{( m_\chi^2+\frac{\mu_\chi^2}{4}-\frac{b_\chi}{t_\beta})}{m_V^2} . \label{c2beta}
  \end{eqnarray}

Because $|c_{2\gamma}|\leq 1$, the eq. (\ref{c2gamma}) results a consequence: $|-(m_\chi^2+\frac{\mu_\chi^2}{4}-\frac{b_\chi}{t_\beta})(1+2s^2_W) +2(m_\rho^2+\frac{\mu^2_\rho}{4}-\frac{b_\rho}{t_\gamma})c^2_{W}|\leq  m^2_W$. But the soft breaking parameters, such as $m^2_{\chi}, m^2_{\rho}, b_{\chi}, b_{\rho}$, should be much larger than $m^2_W$ so these parameters must be degenerate.  In addition, the left hand side of (\ref{c2beta}) have also an upper bound, $|c_{2\beta}|\leq 1$, so does the right hand side.   Because of the hierarchy between two breaking scales
$\mathrm{SU(3)}_L$ and $\mathrm{SU(2)}_L$, $m_W\ll m_V$, the first term in right hand side is suppressed, then we have $|(m_\chi^2+\frac{\mu_\chi^2}{4}-\frac{b_\chi}{t_\beta})|\leq \frac{2c^2_W}{3(1-4 s^2_W)} m_V^2$.
 Hence two quantities $(m_\chi^2+\frac{\mu_\chi^2}{4}-\frac{b_\chi}{t_\beta})$ and
$(m_\rho^2+\frac{\mu^2_\rho}{4}-\frac{b_\rho}{t_\gamma})$ are all in the $\mathrm{SU(3)}_L$
scale.  This leads to a interesting constraint of soft breaking parameters of the SUSYM331:  Although the supersymmetry is spontanously broken before the breaking of  the $SU(3)_L$ symmetry, both the soft parameters and the $SU(3)_L$ breaking scale should be in the same order.  This is very interesting point that did not mention in \cite{Ferr}.  This conclusion also explains why the  values of parameters   $b_{\rho}$ and $b_{\chi}$ are  chosen in \cite{Ferr} in order to get consistent values of the lightest CP-even neutral Higgs   mass.

    Although the Higgs sector of the SUSYRM331 was investigated in \cite{Ferr},  two squared mass matrices of neutral and DCHs are only numerically estimated with some specific values of parameter space. But we think that before starting a numerical calculation it is better to find approximate  expressions of these masses in order to predict the  reasonable ranges of the parameters in the model, as what  we  will represent  through this work. More important, we will show  that approximate expressions are  very useful to determine many interesting properties of the Higgs spectrums.

   The Higgs spectrum are listed as follows
\begin{enumerate}
    \item \textbf{CP-odd neutral Higgses}. Two massless Higgses
    eaten by two neutral gauge bosons are
\bea
 H_{A_3}&=&F_{\chi'}\cos\beta-F_\chi\sin\beta, \hs
 H_{A_4}=F_{\rho'}\cos\gamma-F_\rho\sin\gamma. \label{cpoddmasslessHiggs}
\eea
 Two massive Higgses are expressed in terms of  the original
 Higgses as follows
\[  H_{A_1}=F_{\rho}\cos\gamma+F_{\rho'}\sin\gamma, \hs
 H_{A_2}=F_{\chi}\cos\beta+F_{\chi'}\sin\beta.
\]
 and their masses are
 \be m^2_{A_1}= \frac{2b_{\rho}}{s_{2\gamma}}=m^2_{\rho}+m^2_{\rho'}+\frac{1}{2}\mu^2_{\rho},
  \hs m^2_{A_2}=\frac{2b_{\chi}}{s_{2\beta}}=m^2_{\chi}+m^2_{\chi'}+\frac{1}{2}\mu^2_{\chi}.
  \label{pseudomass2}\ee
\item  \textbf{Singly charged Higgses}.  Two massless eigenstates of
 these Higgses are
\[ H_3^{\pm} =\chi^{\pm}\sin\beta+{\chi'}^\pm\cos\beta, \hs
 H_4^{\pm}=\rho^{\pm}\sin\gamma+{\rho'}^\pm\cos\gamma, \]
which are eaten by the singly charged gauge bosons. Two other massive states are
\bea H_1^{\pm}
&=&-\rho^{\pm}\cos\gamma+{\rho'}^\pm\sin\gamma, \hs m^2_{H^{\pm}_1}= m^2_{A_1}+m^2_W, \\
H_2^{\pm} &=&-\chi^{\pm}\cos\beta+{\chi'}^\pm\sin\beta, \hs
m^2_{H^{\pm}_2}= m^2_{A_2} +m^2_V.\nn
 \eea
\item \textbf{CP-even Neutral Higgses}. In the basis of $(H_{\rho},~ H_{\rho'},~ H_{\chi},~
 H_{\chi'})$ mass term  of the neutral scalar Higgses has form of
 \bea \mathcal{L}_{H^0}&=&\frac{1}{2}( H_{\rho},~ H_{\rho'},~ H_{\chi},~
 H_{\chi'})\times~\mathcal{M}^2_{4H}\times( H_{\rho},~ H_{\rho'},~ H_{\chi},~
 H_{\chi'})^T,
 \label{maphiss1}\eea
  where
\bea   M_{4H}^2=\left(%
\begin{array}{cccc}
    m^2_{S11} & m^2_{S12} & m^2_{S13} & m^2_{S14} \\
              & m^2_{S22} & m^2_{S23} & m^2_{S24} \\
              &           & m^2_{S33} & m^2_{S34} \\
              &           &           & m^2_{S44} \\
\end{array}%
\right).\nn \eea  Analytic formulas of entries of the matrix were
listed in \cite{huong1,Ferr}. This matrix has a problem of finding the exact analytic expressions of eigenvalues, the reason why the ref. \cite{Ferr} had to choose the approach of numerical investigation.

 We remind that the eigenvalues of
this matrix, $\lambda=m^2_{H^0}$,  must satisfy the equation \be
f(\lambda)\equiv\det(M^2_{4H}-\lambda I_4) = 0.\label{nHiggs1}\ee
 As function of $\lambda$,  the left hand side of (\ref{nHiggs1}) is a polynomial of degree 4.  Based on the very detail discussion on  Higgs spectrum of the SUSYE331 in \cite{susye331r} that we do not repeat again,  this function  can be expressed in terms of  independent parameters $m_{A_1}, m_{A_2},c_{2\gamma}, c_{2\beta}, m_W$ and $m_V$ where $m_{A_1}$ and $m_{A_2}$ are soft breaking parameters.  As commented above, these soft parameters are the same orders of the $m_V$-the $SU(3)_L$ scale, i.e $m_{A_1}/m_V, m_{A_2}/m_V\sim \mathcal{O}(1)$.  To find  approximate expressions of Higgs masses, it is needed to define a very small parameter: $\epsilon=\left(m^2_W/m_V^2\right)\leq \left(80.4/2000\right)^2=0.0016$.  Then the masses of these neutral Higgses  can be written as expansions of powers of $\epsilon$,
\bea
 m^2_{H^0_1}&=& M^2_{Z}c^2_{2\gamma}+ \mathcal{O}( m_W^2) \times \epsilon,
   \crn
    m^2_{H^0_2}&=& M^2_{A_1}+ \mathcal{O}( m_W^2),
   \crn
    m^2_{H^0_{3,4}}&=& \frac{1}{6}\left[\frac{4c_W^2m_V^2}{1-4s_W^2}+3m_{A_2}^2\pm
   \sqrt{-\frac{48c_{2\beta}^2m_{A_2}^2m_V^2}{1-4s_W^2}+\left(3m_{A_2}^2+
   \frac{4c_W^2m_V^2}{1-4s_W^2}\right)^2}\right]\crn&+&\mathcal{O}( m_W^2).\label{nHiggs}\eea
   It is necessary to note that the lightest mass with tree-level value of $m_Z|\cos2\gamma|\leq
   m_Z=92$ GeV, being consistent with numerical result shown in \cite{Ferr}.
   Thus, the mass including loop corrections  will
   increase to the current value of 125-126 GeV.
\end{enumerate}
Although the Higgs sector was investigated in \cite{Ferr}, we should emphasize new feature in our work.  To estimate  the tree-level mass of the lightest CP-even Higgs,  by using the reasonable approximation  we have obtained an analytic formula being very consistent with that given in the MSSM. The interesting point is that our result depends only on the condition that all soft parameters must be in the $SU(3)_L$ scale.  The result also  suggests that the $\gamma$ parameter in the SUSYRM331 model plays  the very similar role of the $\beta$ parameter  in the MSSM, defined as the ratio of two VEVs.  This approximation  is very useful  for  estimating masses as well as  predicting many interesting properties of  the DCHs, as we will do in this work.  The authors in \cite{Ferr} also considered only top quark and its superpartner for investigating one-loop corrections to the mass of the lightest neutral Higgs,  then used  this allowed value  to  constrain masses of the DCHs.  But unlike the MSSM,  the  SUSYRM331 contains  new heavy exotic quarks and their superpartners, leading to the fact that  their loop corrections to the mass of the  lightest neutral Higgs have to be considered.  Because the masses of  these new quarks are arbitrary, one cannot tell much about the constraints of charged Higgs masses from considering these  loop corrections.

The approximate formula (\ref{nHiggs}) of neutral Higgs masses is useful
for finding mass eigenstates of the neutral Higgses.
We will consider the two following  rotations for the squared mass matrix
of neutral Higgses appearing in (\ref{maphiss1}),
\bea    C^{\mathrm{n}}_1 &=& \left(
                      \begin{array}{cccc}
                        -c_{\gamma} &s_{\gamma} & 0 & 0 \\
                        s_{\gamma} & c_{\gamma} & 0 & 0 \\
                        0 & 0 & 1 & 0 \\
                        0 & 0 & 0 & 1 \\
                      \end{array}
                    \right), \hs C^{\mathrm{n}}_2 = \left(
                      \begin{array}{cccc}
                       1& & 0 & 0 \\
                        0 & c_{\alpha} & -s_{\alpha} & 0 \\
                        0 & s_{\alpha} &  c_{\alpha} &0 \\
                        0 & 0 & 0 & 1 \\
                      \end{array}
                    \right),
 \label{Nctran1}\eea
 where $s_{\alpha}\equiv\sin\alpha$ and $c_{\alpha}\equiv\cos\alpha$ will
 be defined later.
 Taking the first rotation we obtain
\bea   && C^{n}_1 M^2_{4H}C^{nT}_1 =\crn
&=&\left(
                      \begin{array}{cccc}
                      m^2_{A_1}+\frac{2c^2_{2\gamma}(t^2+2)}{3}m^2_W&\frac{s_{4\gamma}(t^2+2)}{3}m^2_W & \frac{2s_{2\gamma}s_{\beta}(t^2+1)}{3}m_W m_V & - \frac{2s_{2\gamma}c_{\beta}(t^2+1)}{3}m_W m_V \\
                       &\frac{2c^2_{2\gamma}(t^2+2)}{3}m^2_W  &  \frac{2c_{2\gamma}s_{\beta}(t^2+1)}{3}m_W m_V &  -\frac{2c_{2\gamma}c_{\beta}(t^2+1)}{3}m_W m_V \\
                        & & c^2_{\beta}m^2_{A_2}+ \frac{2s^2_{\beta}(t^2+2)}{3}m_V^2& s_{2\beta}\left[ \frac{m^2_{A_2}}{2}+(t^2+2)m^2_V\right] \\
                         &  &  &  s^2_{\beta}m^2_{A_2}+ \frac{2c^2_{\beta}(t^2+2)}{3}m_V^2 \\
                      \end{array}
                    \right).\crn
 \label{Nmassmatrix}\eea
The first diagonal entry of (\ref{Nmassmatrix}) is equal to the
 largest contribution to $m^2_{H_2^0}$ and the sub-matrix including
 entries $\left( C^{n}_1 M^2_{4H}C^{nT}_1\right)_{(i,j=4,5)}$ give two other
 values of heavy masses $m^2_{H^0_3}$ and $m^2_{H^0_4}$.
 While the entry $\left( C^{n}_1 M^2_{4H}C^{nT}_1\right)_{33} = \frac{2c^2_{2\gamma}
 (t^2+2)}{3}m^2_W=\mathcal{O}(m^2_W)$ relates with the lightest Higgs mass,
 but it is different from that shown in (\ref{nHiggs}). To get a right  value
 it must give more contributions from  non-diagonal tries containing factors
  $m_Wm_V$ after taking other rotations. As mentioned, the most interesting values
  of $\alpha$ and $\beta$ satisfy $c_{2\gamma},c_{2\beta}\rightarrow -1$,
  i.e $s_{\gamma,\beta}\rightarrow1$ and $c_{\gamma,\beta}\rightarrow0$.
  It suggests that the largest correction to the lightest mass is from the entries
  $\left( C^{n}_1 M^2_{4H}C^{nT}_1\right)_{23}$ and
   $\left( C^{n}_1 M^2_{4H}C^{nT}_1\right)_{32}$.
  So taking the second rotation with $C^{\mathrm{n}}_2$
 given in (\ref{Nctran1}) and using the limits
  $m_W^2\ll m^2_{V}$ and $c_{\beta}\rightarrow 0$, it is easy to confirm
  that $\left(  C^{n}_2 C^{n}_1 M^2_{4H}(C^{n}_2C^{n}_1)^T\right)_{(22)}\simeq m^2_{Z}|c_{2\gamma}|$
  with $\alpha$ determined by
\be \tan{2\alpha}\equiv \frac{ 4 c_{2\gamma}s_{\beta}(t^2+1)m_Wm_V}{3c^3_{\beta}m^2_{A_2}+2 s^2_{\beta}(t^2+2)m^2_V-2c^2_{2\gamma}(t^2+2)m^2_{W}}\sim \mathcal{O}(m_W/mV). \label{t2al}\ee
Then we can estimate the contribution of the original Higgs states to the mass eigenstates of the even-CP neutral  Higgses $H^0_1$ and $H^0_2$  are
\be  H_{\rho}\rightarrow c_{\alpha} s_{\gamma} H^0_1-c_{\gamma}H^0_{2}, \hs  H_{\rho'}\rightarrow c_{\alpha} c_{\gamma} H^0_1+ s_{\gamma}H^0_{2}, \hs H_{\chi}, H_{\chi'}\rightarrow \mathcal{O}(s_{\alpha}) H^0_1. \label{LnHiggst}\ee
It is interesting that in the decoupling regime where the SUSY and $SU(3)_L$
scales are much larger than the $SU(2)_L$ scale, we have $H_{\rho}\simeq  s_{\gamma} H^0_1
-c_{\gamma}H^0_{2}$ and $H_{\rho'} \simeq c_{\gamma} H^0_1+ s_{\gamma}H^0_{2}$,
 the same as those given in the MSSM\cite{adjo}. Therefore all couplings of these
 two Higgses with  $W^{\pm}$ and $Z$ bosons are the same as those in
 the MSSM. More interesting the SUSYRM331 contains a set of Higgses
  including $m_{H^0_{1,2}},~m_{A_1}$ and $m_{H^{\pm}_1}$  that has similar properties
  to the Higgs spectrum of the MSSM. This property of the SUSY versions of the 3-3-1 models
   was also indicated  previously \cite{susye331r}. As the result, the SUSY Higgs
   contributions to the $\rho$ parameter are the same in both MSSM and SUSYRM331 in
   the decoupling regime.

At the tree level, the above analysis indicates that  the Higgs spectrum can be determined by unknown
 independent parameters: $\gamma,~\beta~, m_V,~m_{A_1}$ and $m_{A_2}$.
Furthermore, the squared mass matrices of both CP-even neutral and DCHs  depend  explicitly
on $c_{2\beta},~s_{2\beta},~c_{2\gamma}$ and $s_{2\gamma}$  but not $t_{2\gamma}$, $t_{2\beta}$. Hence it can be guessed that  Higgs masses will not increase  to infinities when
$t_{\gamma}$ and $t_{\beta}$ are very large. And in some cases we can take the limits
$c_{2\beta,2\gamma}\rightarrow -1$ and $s_{2\beta,2\gamma}\rightarrow 0$ without any inconsistent
calculations. In addition, we will
use the limit $2 \mathrm{TeV}\leq m_V \leq 3$ TeV based on the latest update discussed above. The $m_{A_1}$
and $m_{A_2}$ are the same order of $m_V$ so we set $m_{A_1},~m_{A_2}\geq 1$ TeV
in our calculation.  Other well known values will be used are mass of the $W$ boson $m_W=80.4$ GeV, sine of
the Weinberg angle $s_W=0.231$, mass and total decay width of the $Z$ boson $m_Z= 91.2$ GeV,
$\Gamma_Z=2.46$ GeV. The discovery of the lightest CP-even Higgs mass
of 125 GeV implies that $|c_{2\gamma}| \simeq 1$, i.e $t_{\gamma}$
should be large enough, similar to the case of MSSM. As a consequence,
the relation (\ref{c2gamma}) shows the fine tuning among soft parameters
and the relation (\ref{c2beta}) predicts that $|c_{2\beta}|$ should also be large.
 Therefore  we will fix $t_{\beta}=5$ and $t_{\gamma}=10$ in the numerical investigation which can be applied for the general case of large $t_{\beta}$ and
$t_{\gamma}$. This can be understood  from the reason that all quantities we consider below depend
on $\gamma$, $2\gamma$, $\beta$  and $2\beta$ only by factors of sine or cosine but not tan
functions.

 \section{\label{secdch}Doubly charged Higgs bosons  and  couplings}

 \subsection{ Mass spectrum and properties of the lightest DCH}
Consider the DCHs, the SUSYE331 model contains 8  degrees of freedom after final symmetry breaking. Therefore the squared mass matrix is $4\times4$ and we cannot find the exact expressions for the physical masses. We will treat them the same as the case of  the neutral CP-even Higgses, in much more details to discover all possible interesting properties of the DCHs, especially the lightest.

 The mass term of the doubly charged boson is:
\bea \mathcal{L}_{H^{\pm\pm}}= \left(%
\begin{array}{cccc}
  \rho^{++}, & \rho^{\prime++}, &\chi^{++}, & \chi^{\prime++} \\
\end{array}%
\right) \mathcal{M}^2_{H^{\pm\pm}} \left(%
\begin{array}{cccc}
  \rho^{--} & \rho^{\prime--} &\chi^{--} & \chi^{\prime--} \\
\end{array}%
\right)^T, \nn  \eea
 where the elements of the squared mass  matrix was shown precisely
 in \cite{Ferr}. Taking a rotation characterized by a matrix
 \be  C_1 = \left(%
\begin{array}{cccc}
  \frac{-m_W s_{\gamma}}{m_U} & 0 & c_{\gamma} & \frac{m_V s_{\gamma}}{m_U} \\
   \frac{-m_W c_{\gamma}}{m_U} & 0 & -s_{\gamma} & \frac{m_V c_{\gamma}}{m_U}\\
   \frac{m_V s_{\beta}}{m_U} & -c_{\beta} & 0 &\frac{m_W s_{\beta}}{m_U} \\
   \frac{m_Vc_{\beta}}{m_U} & s_{\beta} & 0 & \frac{m_W c_{\beta}}{m_U} \\
\end{array}%
\right)\label{rotation1} \ee
 we get  new squared mass matrix
{\small \bea \mathcal{M}^2_{H^{\prime\pm\pm}}\equiv C_1^T \mathcal{M}^2_{H^{\pm\pm}}C_1= \left(%
\begin{array}{cccc}
  0 & 0 & 0 & 0 \\
  0 & m^2_{A_2}+m^2_V-c_{2\beta}c_{2\gamma}m^2_W & -s_{2\beta} s_{2\gamma}m_V
   m_W  &c_{2\gamma}s_{2\beta} m_U m_W \\
   0 &  -s_{2\beta} s_{2\gamma}m_V m_W  & m_{A_1}^2 - c_{2\beta} c_{2\gamma} m_V^2 +
    m_W^2 &  -s_{2\gamma}
   c_{2\beta}m_U m_V  \\
  0 & c_{2\gamma}s_{2\beta} m_U m_W &  -s_{2\gamma}c_{2\beta}m_U m_V &
   c_{2\gamma}c_{2\beta}m^2_U \\
   \end{array}%
\right).\crn \label{massd1}\eea}
 Corresponding to massless solution in  (\ref{massd1}), the Goldstone  eaten by  the  doubly charged gauge boson  is
represented exactly in term of original Higgses,
 \be G^{\pm\pm}= -\frac{m_W s_{\gamma}}{m_U} \rho^{\pm\pm}-\frac{m_W c_{\gamma}}{m_U}
 \rho^{\prime\pm\pm}+\frac{m_Vs_{\beta}}{m_U} \chi^{\pm\pm}+\frac{m_V c_{\beta}}{m_U}
 \chi^{\prime\pm\pm}.
 \label{dgoston}\ee
Because $m_W\ll m_U$, $m_V\simeq m_U$ and $s_{\gamma}\leq 1$, the
doubly charged gauge boson couples weakly to light Higgses but strongly to heavy Higgses.

The squared mass matrix of the DCHs  (\ref{massd1}) also shows that if
there exist a light DCH (i.e $ \sim \mathcal{O}(m^2_W)$) then the
contributions of the off-diagonal entries to the mass of this Higgs are large. Then it is
difficult to find an  analytic formula for both mass eigenstates and eigenvalues.  Note that apart from Goldstone boson (\ref{dgoston}), there are three other states denoted by $H'^{\pm\pm}_i, ~(i=1,2,3)$. They relate to original DCHs by a transformation
\bea  \left(%
\begin{array}{cccc}
  \rho^{\pm\pm}, & \rho^{\prime\pm\pm}, & \chi^{\pm\pm}, & \chi^{\prime\pm\pm} \\
\end{array}%
\right)^T=C_1\left(%
\begin{array}{cccc}
  G^{\pm\pm}, &  H^{\prime\pm\pm}_1,& H^{\prime\pm\pm}_2, & H^{\prime\pm\pm}_3 \\
\end{array}%
\right)^T.  \label{medchargedHiggs}\eea
  We assume that three physical  DCHs  relate to
$(H^{\prime\pm\pm}_1$,   $ H^{\prime\pm\pm}_2$,  $
 H^{\prime\pm\pm}_3 )$ by a $3\times 3$ matrix
$\Lambda$ as follows
 \be \left(H^{\prime\pm\pm}_1,\;H^{\prime\pm\pm}_2,\;H^{\prime\pm\pm}_3
\right)^T= \Lambda
\left(H^{\pm\pm}_1,\;H^{\pm\pm}_2,\;H^{\pm\pm}_3 \right)^T.
\label{machcsdhiggs}\ee
To estimate the values of entries of the matrix $\Lambda$, we firstly find out some properties of
mass eigenvalues of  the DCHs.  The  remain three eigenvalues of this matrix
$\lambda=m_{H^{\pm\pm}}^{2}$ must satisfy  the equation $\det(\mathcal{M}^2_{H^{\pm\pm}} -
\lambda I_4) = 0$, or equivalently, $\lambda f(\lambda)=0$ with
 \be
 f(\lambda)=a\lambda^3+b\lambda^2+c\lambda+d,\label{fdoublycHiggs} \ee
 where
\bea
a&=&-(m^2_V +  m^2_{A_1} +m^2_{A_2}+m^2_W),  \crn
b&=&-c_{2\beta}^2 m^4_{V}+ m^2_{A_1}\left( m^2_V+c_{2\beta} c_{2\gamma}+ m^2_{A_2}
\right)+\left[m^2_{A_2} + c_{2\beta} c_{2\gamma} (2 m^2_V + m^2_{A_2})\right]m^2_W -c_{2\gamma}^2
m^4_W, \crn
 c&=& \left(m^2_V +m^2_W\right)\left[c_{2\beta} m^2_V- c_{2\gamma} (m^2_W
 +m^2_{A_1})\right]
 \left[c_{2\beta}(m^2_V+m^2_{A_2}) - c_{2\gamma}m^2_W\right].\label{factordcHiggs} \eea
 This equation gives three solutions corresponding to three  masses
  of  the physical DCHs at tree level.  We denote them as $m^2_{H^{\pm\pm}_i}$ with  $i=1,2,3$ and $m^2_{H^{\pm\pm}_1}, ~m^2_{H^{\pm\pm}_2}>m^2_{H^{\pm\pm}_3}$. Combining the last equation of (\ref{factordcHiggs}) with the vieta's formula,   in order to avoid  the appearance of  tachyons  we deduce  that
 \be m_{H^{\pm\pm}_1}^{2}m_{H^{\pm\pm}_2}^{2}m_{H^{\pm\pm}_3}^{2}
  =-c>0\Leftrightarrow \frac{\left(m^2_{A_1}+m^2_W\right) c_{2\gamma}}{m^2_V}
 <c_{2\beta}<\frac{m^2_W  c_{2\gamma}}{m^2_V+m_{A_2}^2}<0. \label{ctachyon}\ee
Furthermore, the entry $ ()_{22}$ of (\ref{massd1}) suggests that  if $ m^2_{A_1}$ is enough close to $m^2_V$ then there may appear
one light DCH, while two other values are  always in the $\mathrm{SU(3)}_L$ scale. So in order to
 to find the best approximate formulas of vertex factors $V^0H^{++}H^{--}$, it is better to  investigate the mass values of the DCHs using
 techniques shown in \cite{susye331r}, and partly mentioned when discussing on the neutral CP-even Higgs sector.  The masses can be
 expanded as
\be  m^2_{H^{\pm \pm}}=
  X' m^2_V + X''\times m^2_W+ \mathcal{O}(\epsilon)\times m^2_W. \label{n0hvh1} \ee
   The heavy Higgses satisfy the condition $X' \sim\mathcal{O}(1)$, i.e in the soft-breaking or $SU(3)_L$ scale.  Keeping only the leading term in (\ref{n0hvh1}) as the  largest contribution,  the  masses of the three DCHs  are
\bea m^2_{H^{\pm \pm}_1} &\simeq & m^2_V+m^2_{A_2},
 \crn m^2_{H^{\pm \pm}_{2,3}} &\simeq &
\frac{1}{2}\left(m^2_{A_1} \pm  \sqrt{4 c_{2\beta}^2 m^4_V - 4 c_{2\beta} c_{2\gamma}m^2_V
m^2_{A_1} +m^4_{A_1}} \right). \label{dchmass1} \eea
 Comparing the $\left(\mathcal{M}^2_{H^{\prime\pm\pm}}\right)_{33}$ entry of (\ref{massd1}) with
$m^2_{H^{\pm \pm}_1}$ in the first line of (\ref{dchmass1}), it can be realized that
$\left(\mathcal{M}^2_{H^{\prime\pm\pm}}\right)_{33}-m^2_{H^{\pm \pm}_1}= \mathcal{O}(m^2_W) \ll
m^2_{H^{\pm \pm}_1}$. As a result, the main contribution to the mass eigenstate of
$m^2_{H^{\pm\pm}_1}$ is $H^{\prime\pm\pm}_1=-c_{\beta}\chi^{\pm\pm}+s_{\beta}\chi^{\prime\pm\pm}$, i.e $\Lambda_{1i},\Lambda_{i1}\simeq \delta_{1i}$
with $i=1,2,3$. This is very useful to find the formulas of coupling between these DCHs
with neutral gauge bosons.

\begin{figure}[h]
  \centering
\begin{tabular}{cc}
\epsfig{file=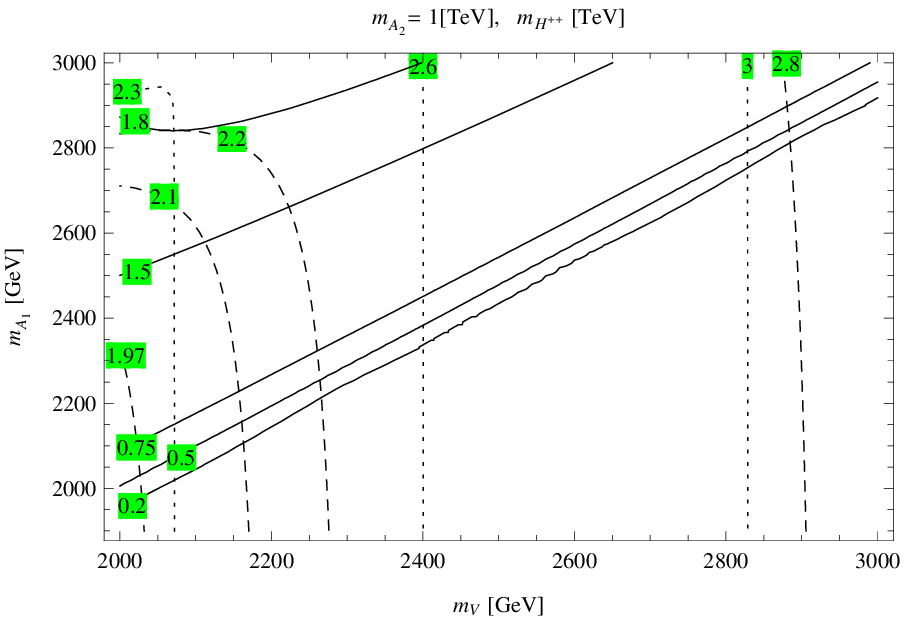,width=0.5\linewidth,clip=}
&
\epsfig{file=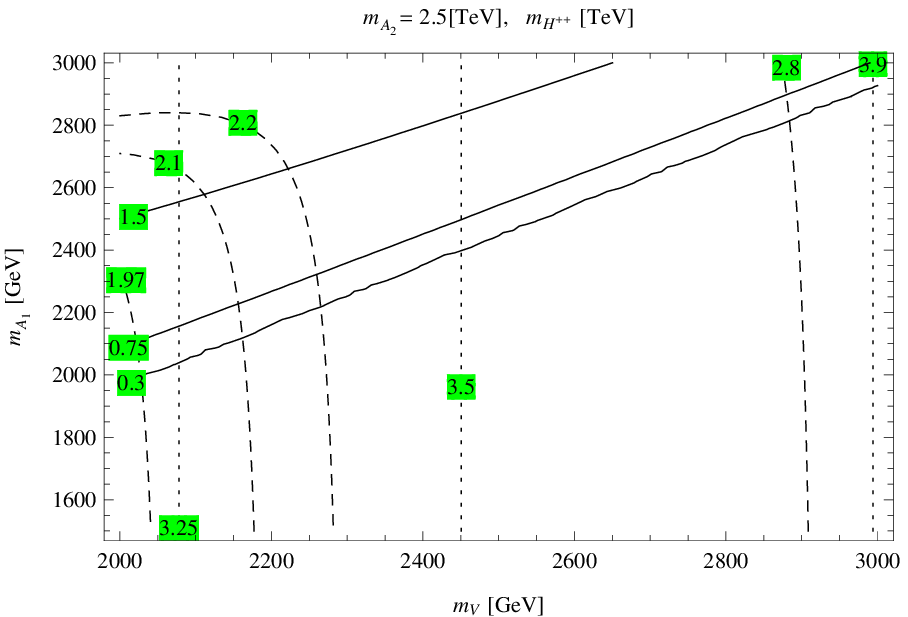,width=0.5\linewidth,clip=}
\\
\end{tabular}
  \caption{Contour plots for masses of  DCHs as functions of $m_{A_1}$ and $m_V$.
    The left (right) panel corresponds to  $m_{A_2}=1$ (2.5) TeV. Here the heaviest: dotted curves,
    the second heavy:
    dashed curves, the lightest: the black curves.}
   \label{ccHmass1}
\end{figure}

 The above approximative formulas of   DCH masses can predict precisely constraints of   these masses.  From eq. (\ref{fdoublycHiggs}),  applying  Vieta's formulas to  the first line of (\ref{factordcHiggs}) we get a relation
$$m^2_{H^{\pm\pm}_1}+m^2_{H^{\pm\pm}_2}+m^2_{H^{\pm\pm}_3}=
m^2_V+m^2_{A_2}+m^2_{A_1}+m^2_{W}.$$
 Combining with $m^2_{H^{\pm\pm}_1}=m^2_V+m^2_{A_2}+\mathcal{O}(m^2_W)$ we have a sum of two DCHs,
$m^2_{H^{\pm\pm}_2}+m^2_{H^{\pm\pm}_3}=m^2_{A_{1}}+\mathcal{O}(m^2_W)$,  which is still in order of
$\mathrm{SU(3)}_L$ scale. So there must be at most one light DCH  in the model. If the
model contains this light Higgs, i.e $m^2_{H^{\pm\pm}_3}\sim\mathcal{O}(m^2_W)$ we can prove that
$m^2_{H^{\pm\pm}_2}\sim m^2_V$. This is the consequence deduced from  the last eq. of (\ref{factordcHiggs}) and (\ref{ctachyon}):
the condition of existing this light Higgs is
\be 0<k_{H^{\pm\pm}}\equiv -c_{2\beta}\left[c_{2\beta} m^2_V-
c_{2\gamma} (m^2_W
 +m^2_{A_1})\right] \sim \mathcal{O}(m^2_W).\label{conditonlhiggs1}\ee
 It leads to $m^2_{H^{\pm\pm}_2}+m^2_{H^{\pm\pm}_3}\sim \mathcal{O}(m^2_V)$. The  mass of
the lightest  DCH $m^2_{H^{\pm\pm}_3}$ depends directly on  the scale of
$k_{H^{\pm\pm}}$.  It is easy to realize that two mass eigenstates  $H^{\pm\pm}_{2,3}$ get main
contributions from $\rho^{\pm\pm}$ and $\rho^{\prime\pm\pm}$. This is consistent with the fact that
main contribution to mass of the heavy Higgs $H^{\pm\pm}_2$ is from the
 $b_{\rho}\rho\rho'$ term. Numerical values of these  masses are
 illustrated in the figure \ref{ccHmass1}.

The condition (\ref{conditonlhiggs1}) give a lower bound of $m_{A_1}>1.8$ TeV. With $m_{A_2}<3$TeV, the heaviest DCH is
always $H^{\pm\pm}_1$, except the case of the very light $m_V\simeq 2$ TeV.
This explains why $m_{H^{\pm\pm}_1}$ does not depend on $m_{A_1}$,
while it is sensitive with $m_V$. The second heavy DCH is also independent
with small value of $m_{A_1}$, which can be explained as follows.
The condition of avoiding tachyon DCH (\ref{ctachyon}) implies
that  $0<c_{2\beta}/c_{2\gamma} m_V^2-m^2_W < m_{A_1}^2$.
Therefore, the small  $m_{A_1}$ will give
$c_{2\beta} m_V^2- c_{2\gamma}m_{A_1}^2 \sim \mathcal{O}(m_W^2)$
which is the condition for appearing the very light DCH. This gives
$$ \frac{4c_{2\beta} m_V^2(c_{2\beta} m^2_V -  c_{2\gamma}m^2_{A_1})}{m^4_{A_1}}\sim \frac{m_V^2}{m^2_{A_1}}\times \frac{\mathcal{O}(m_W^2)}{m^2_{A_1}}\ll 1 $$
which we can use for estimating an approximation of $m^2_{H^{\pm\pm}_2}$
$$ m^2_{H^{\pm\pm}_2}\simeq \frac{\left(c_{2\beta}m_V^2-c_{2\gamma}m^2_{A_1}\right)^2}{m^2_{A_1}}+ s^2_{2\gamma}m^2_{A_1}+c_{2\beta}c_{2\gamma}m_V^2.$$
 Here the very small $s^2_{2\gamma}$ is assumed in this work.  This
 means that $m^2_{H^{\pm\pm}_2}$ is sensitive with the changes of
 $c_{2\beta}c_{2\gamma}m_V^2$ but not with the small changes of $m_{A_1}$.

Now we  pay attention to the first interesting property relating to the SUSYRM331:
 it may contain the
lightest DCH that does not depend on the $SU(3)_L$ scale but the specific correlation
between  $m_{A_1}$ and $m_V$,  as indicated in (\ref{conditonlhiggs1}) and illustrated
in the figure
\ref{ccHmass1}. It can be seen that there always exists a region of the parameter space
containing
the mass of this DCH with order of $\mathcal{O}(100) $ GeV. So the ILC can create this
Higgs  in the   collision
energy of 0.5-1.0 TeV.
 While the two other DCHs are very heavy because of the large lower  bound of
 $m_V\geq2$ TeV as well as $m_{A_1}>1.8$ TeV obtained from the condition (\ref{conditonlhiggs1}).  The lower bound of the the first DCH mass is
 $m_{H^{\pm\pm}_1}\simeq \sqrt{m_V^2+m^2_{A_2}}=m_{H^{\pm}_2}> 2$TeV.
 The additional condition of $m_{A_1}>1.8$TeV will result a larger lower bound of
 $m_{H^{\pm\pm}_1}>3$ TeV and be  independent with  $m_{A_2}$. The lower bound of
 $m^2_{H^{\pm\pm}_2}$ directly depends  on the condition  (\ref{conditonlhiggs1}),
 where $m^2_{A_1}>\frac{c_{2\beta}m^2_V}{c_{2\gamma}}-m^2_W$, leading to
 $ m_{H^{\pm\pm}_2}>1.9$ TeV, when $c_{2\beta},c_{2\gamma}\simeq -1$  are  assumed
 in this work.  Hence the SUSYRM331 model predicts that the DCHs will not appear in
 the $e^+e^-$ colliders with colliding energies below 4 TeV,  except the lightest.
 \begin{figure}[h]
  \centering
\begin{tabular}{cc}
\epsfig{file=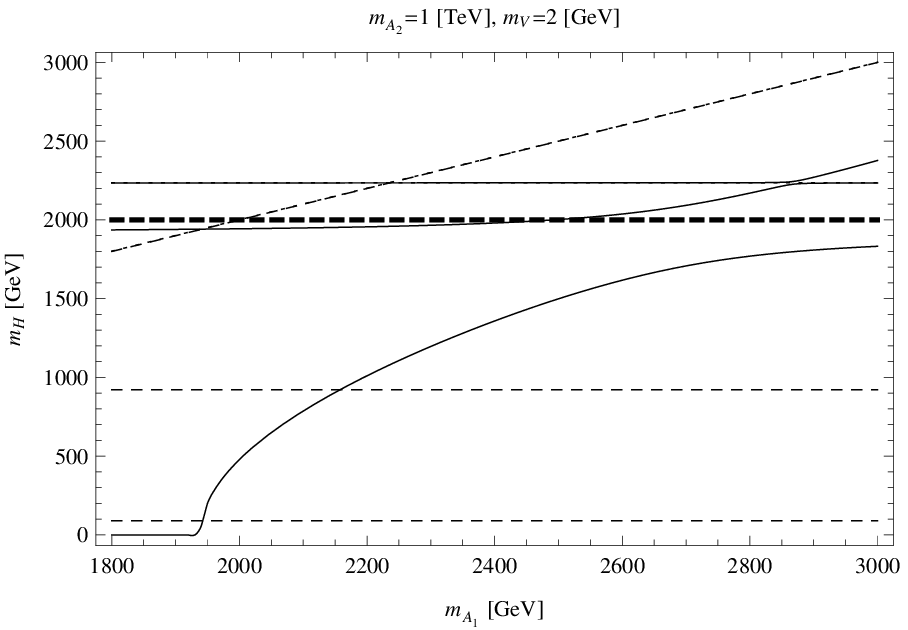,width=0.5\linewidth,clip=} &
\epsfig{file=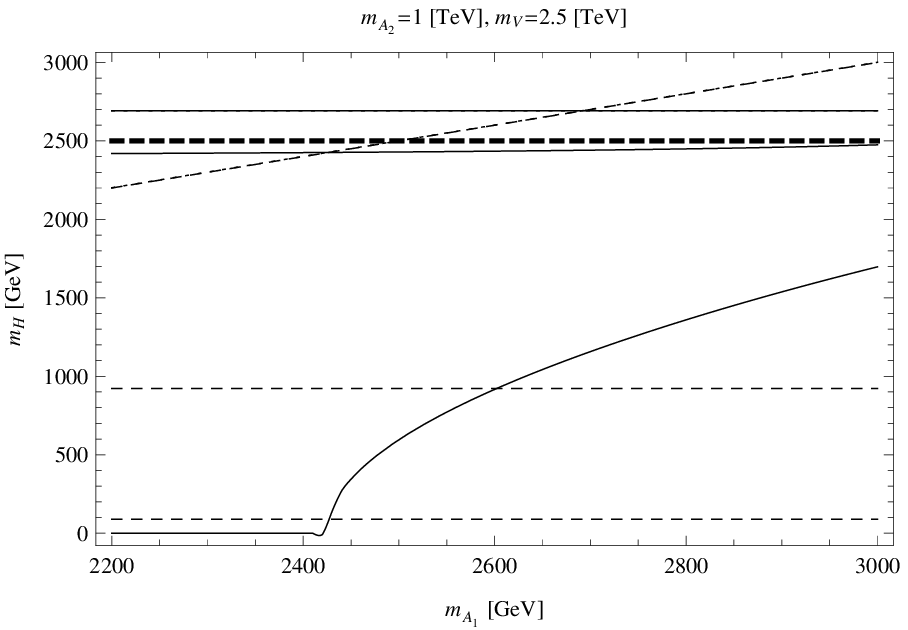,width=0.5\linewidth,clip=}
\\
\end{tabular}
  \caption{Plots of  mass spectrum  as functions of $m_{A_1}$ with different fixed $m_V$. The black,
     dotted,  dashed and  thick dashed curves represent
     DCHs, singly charged, neutral Higgses and $V$ gauge boson, respectively.}
   \label{massspec1}
\end{figure}

There is the second interesting property of the lightets DCH: it is
lighter than all particles including new gauge bosons, singly
charged Higgses, as illustrated in the  figure
\ref{massspec1}.  It is easy to see this when we compare all masses  computed above. The exotic quarks as well as superpartners can be  reasonably supposed to be heavier  than the lightest DCH,  then we do not consider here.  Then we can indicate that  the lightest DCH  decays into only pair of charged leptons.  Recall that  all DCHs have
  the lepton number two. Therefore the total lepton number of all final states of their decays must be  the same. In particularly,  the final states  of each decay should contain one bilepton or a pair of charged leptons. From the table \ref{hcc3coupling} and \ref{hcc4coupling} that list all three- and four-vertex couplings relating with DCHs, we can see that:  except the coupling with two leptons,  the lightest DCH always couples with at least
 one heavier particle: another DCH, a singly charged Higgs, a new gauge boson or a CP-odd neutral
 Higgs.   So if  existing the most promising  signal  of  the lightest DCH  is the decay into only  a pair of leptons.  This strongly suggests the possibility of detection  of the lightest DCH in $e^+e^-$ colliders such as the ILC or CLIC, even in the low  energy of 0.5-1 TeV.

  As mentioned above, the  state of the lightest DCH is contributed mainly from two Higgses $\rho$ and $\rho'$. Combining this with coupling factors     between DCHs and charged leptons shown in the table \ref{hcc3coupling},  it is easy to prove that
    the partial decay of this DCH to a  pair of same-sign leptons is $\Gamma(H^{\pm\pm}\rightarrow l^{\pm}_il^{\pm}_i)
    \sim \left(m_{l_i}/m_W\right)^2$ with $l_i=e,~\mu,~\tau$.  As a result, we obtain
    $\mathrm{Br}(H^{\pm\pm}\rightarrow \tau^{\pm}\tau^{\pm})\simeq 1$,  i.e, the  number of events of
    four-tauon signals   is  equal to that of creating the lightest DCH at $e^+e^-$ colliders.

Because the lightest DCH mainly decays to the same-sign $\tau$ pairs, the
lower bound from experimental searches is 204 GeV\cite{H2Ex}. This lightest DCH
is very different from other DCHs predicted by other models where they can mainly
decay to pairs of the  two same-sign $W$ bosons or $W^-H^-$ \cite{hww1,hww2,hwh1}.
On the other hand, the heavy DCHs predicted by the SUSYRM331 only couple to  other
 bileptons in the model, including $H^{\pm}_2$, $V^{\pm}$, $U^{\pm\pm}$ and the
 corresponding superpartners.  The most interesting  couplings is
 $H_1^{\pm\pm}W^{\mp}H^{\mp}_2$,  which was discussed
 in \cite{hww1,hww2,hwh1,lrDCH} for creating DCHs at LHC through
 virtual $W^{\pm}$ bosons.  While there are not any couplings
 of  $H^{\pm\pm}W^{\mp}W^{\mp}$ because of the lepton number conversation.
 In addition, the masses of the two heavy DCHs are always larger than 1.5 TeV, they
 do not appear in the $e^+e^-$ colliders such as ILC and CLIC with their recent
 designs. These two Higgses may only appear at the LHC with high luminosity.
 In addition, both of them can be created through the channel of
 $pp\rightarrow \gamma/Z/Z'\rightarrow H^{++}H^{--}$,
 but only $H^{\pm\pm}_1$
 may be created through the channel $pp\rightarrow W^{\pm}\rightarrow H^{\pm\pm}_1H^{\mp}_2$.
 Regarding to the latter channel, discussions in refs. \cite{hwh1,lrDCH}
 indicated that it is very hard to find signals of these very heavy DCH,
 even at the very high luminosity of $3000~\mathrm{fb^{-1}}$ that LHC can reach.
 While the former happens for all three DCH, the two heavy DCHs are also very
 hard to observed \cite{lrDCH1}.

From the above reason, the SUSYRM331 predicts that only the lightest
DCH may be discovered at  the LHC and $e^+e^-$ colliders and the signal can be observed
through the main channel of
$pp/e^+e^-\rightarrow \gamma /Z/Z'\rightarrow H^{++}_3H^{--}_3 \rightarrow $
four tauons. With the LHC, one hopes it will be observed up to mass of 600 GeV with high
luminosity of $3000 \mathrm{fb^{-1}}$. Because the creating cross section is proportional to $1/s^2$, with $s$ being colliding energy, the signal of DCH at ILC and CLIC seems better than that in LHC. In addition, with  the ILC or CLIC  a larger range mass of
the DCH can be observed, so we will mainly pay attention to the lightest DCH at
$e^+e^-$ colliders.

Now we will  estimate the
 allowed kinetic condition $2 m_{H^{\pm\pm}}\leq E_{cm}$ for the creation of
 the lightest DCH at the $e^+e^-$
 colliders with maximal center mass (CM) energy of 3 TeV.
Even in case of both large values of $m_V$ and $m_{A_1}$, there always exists a
region where mass of the lightest DCH is in the order of $\mathcal{O}(100)$ GeV.
Furthermore, this
light value almost does not depend on $m_{A_2}$. Although the mass values below 204 GeV of this
Higgs are almost excluded recently from its decay into only a pair of tauons \cite{H2Ex,H2Ex2}, higher values can be searched by ILC  or CLIC with
CM energy about 1 TeV.

The appearance of the light DCH  may give large loop corrections to the decays of well-known particles. The most important is the  decay channel of the SM-like Higgs $H^0_1\rightarrow \gamma\gamma$, which get contributions from only pure loop corrections. The signal strength of this decay is defined as  the ratio of the observed cross section  and the SM predicion,  $\mu_{\gamma\gamma}=  \sigma^{\mathrm{obs}}_{H\rightarrow\gamma\gamma}/\sigma^{\mathrm{SM}}_{H\rightarrow\gamma\gamma}$, and was found to be  slightly excess than 1 \cite{h2ga1,h2ga2,h2ga3,h2ga4}. The enhancement is explained by the contributions of new particles to the partial decay width of $H^0_1\rightarrow \gamma\gamma$ \cite{adjo}. The analytic formula of this decay  width  is the sum of the three particular parts:  SM, $SU(3)_L$ \cite{h2ga331} and SUSY contributions.  The SM and SUSYRM331 contributions can be deduced  based on \cite{adjo,adjo1}.

\begin{figure}[h]
  \centering
  \includegraphics[width=12cm]{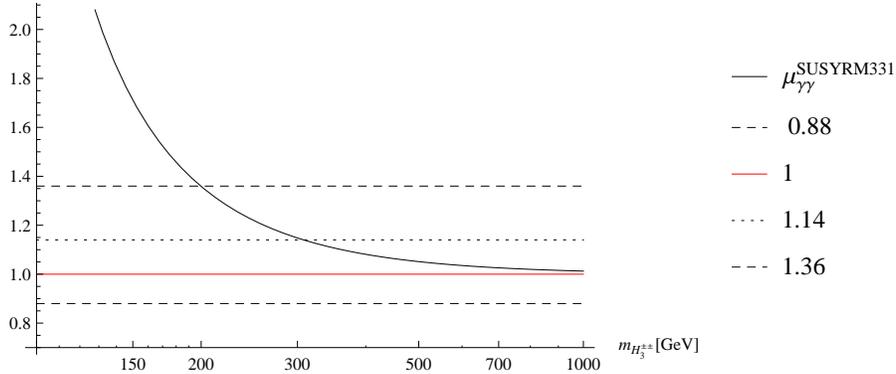}\\
  \caption{ Signal strength of the decay $H^0_1\rightarrow \gamma\gamma$ as function of the lightest DCH mass.}\label{h2ga1}
\end{figure}

The SUSYRM331, with both $SU(3)_L$ and SUSY breaking scales being larger than 7 TeV, results a consequence that most  of the  $SU(3)_L$ and SUSY  particles give suppressed contributions to this decay, except the lightest DCH.  Hence the $H^0_1\rightarrow \gamma\gamma$ is an important channel to set a lower bound to its mass. We will follow the latest update of $\mu_{\gamma\gamma}$  in ref. \cite{h2ga2} where $ \mu_{\gamma\gamma}=1.12\pm 0.24$ without any inconsistences with results of ATLAS. In addition, to simplify the calculation, we consider  the largest new physics effect to the  $H^0_1$ decay is from only the lightest DCH $H^{\pm\pm}_3$ to the partial decay $H^0_1\rightarrow \gamma\gamma$. As a result, we have a very simple formula,
 which must satisfy the experimental constraint: $ 0.88=1.12-0.24\leq\mu^{\mathrm{SUSYRM331}}_{\gamma\gamma}\leq 1.12+0.24=1.36$. The partial decay of the $H^0_1\rightarrow \gamma\gamma$ is written as
\bea  \Gamma^{\mathrm{SUSYRM331}}_{H^0_1\rightarrow \gamma\gamma} &\simeq& \frac{G_{\mu}\alpha^2 m^3_{H^0_1}}{128\sqrt{2}\pi^3}\left| A^{\mathrm{SM}}+ \Delta A\right|^2,\label{h2gssusy1} \eea
where $A^{\mathrm{SM}}$ is the contribution from the SM particles, and $\Delta A$  is the new contribution from SUSYRM331 particles. The well-known SM formula can be found in many textbooks, for example in \cite{adjo1}.  To find a simple analytic formula,  our work  considers  only case of $c_{2\gamma},c_{2\beta}\rightarrow -1$, where  the masses of the DCHs are nearly equal to the diagonal entries of the squared mass matrix (\ref{massd1}), being consistent with (\ref{dchmass1}).  The lightest DCH mass  now satisfies $m^2_{H^{\pm\pm}_3} = \mathcal{O}(100)$  GeV when (\ref{conditonlhiggs1}) is satisfied. And also, the main contribution to the mixing matrix of the DCHs is  $C_1$ shown in (\ref{rotation1}). Combining with discussion on the neutral Higgs sector, we get the $H^0_1H^{++}_3H^{--}_3$ coupling is $g_{H^0HH}\simeq \frac{1}{3}g(t^2+2)c_{2\gamma}m_W\simeq-\frac{1}{3}g(t^2+2)m_W =- \frac{2c^2_W }{3(1-4s^2_W)}m_W$. Following this, the formula of  $\Delta A$ can be written as\cite{adjo},
\be \Delta A = -\frac{8 c^2_W m^2_W}{3(1-4s^2_W)m^2_{H^{\pm\pm}_3}}A_0(t_H),\label{hpp2ga}\ee
where $t_H=\frac{m^2_{H^0_1}}{4m^2_{H^{\pm\pm}_3}}$ and
\bea
A_0(t)&=&-[t-f(t)]t^{-2},\crn
f(t)&=&\left\{
              \begin{array}{cc}
                \arcsin^2\sqrt{t}  & \mathrm{for}~t\leq1 \\
                -\frac{1}{4}\left[\ln \left( \frac{1+\sqrt{1-t^{-1}}}{1-\sqrt{1-t^{-1}}}\right)- i\pi\right]^2 & \mathrm{for}~t>1. \\
              \end{array}
           \right. \nn\eea
 The signal strength of the decay $H^0_1\rightarrow \gamma\gamma$ predicted  by the SUSYRM331 is shown in the figure \ref{h2ga1}, where
  $m_{H^{\pm\pm}_3}\geq 200$ GeV is allowed, being equal to the lower bound of 200 GeV
  from the current experiments.

 Finally, in order to calculate the cross section of the DCHs in the $e^+e^-$ colliders,  the following of this section  will calculate the coupling of
$H^{++}H^{--}V^0$.

\subsection{Couplings between DCHs with neutral scalars and gauge bosons}

It is noted that the process $e^{+}e^{-}\rightarrow H^{++}H^{--}$
  through virtual neutral Higgses  involves  with the coupling $e^{+}e^{-}H^0$. In  the SUSY version \cite{ffv1},
    this kind of the coupling is $2 g m_e/(m_Wc_{\gamma})$.
    While  the SUSY version \cite{huong1} has no this kind of coupling at the tree level. In this
    work we will use the case in \cite{ffv1}.
      Corresponding with this,     we consider the coupling $H^{++}H^{--}\rho^{\prime0}$ . Couplings $H^{++}H^{--}H^0$  comes
      from the $D$-term of the  scalar potential (\ref{ep1}) , namely
    \bea  \mathcal{L}_{H^{++}H^{--}H^0}&=& \frac{gm_W}{6}
     \rho^{\prime0} \left(
     \begin{array}{cccc}
     \rho^{--}, & \rho^{\prime--}, &\chi^{--}, & \chi^{\prime--} \\
      \end{array}
      \right)\crn
       &\times&  \left(
      \begin{array}{cccc}
      2s_{\gamma}(t^2-1) & -3s_{\gamma} & 0 & 0 \\
              -3c_{\gamma}& 2 c_{\gamma}(t^2+2)& 0 & 0 \\
                 0  &0&  2 c_{\gamma}(t^2-2) & 0 \\
                    0 & 0 & 0 &   -2 c_{\gamma}(t^2-2) \\
         \end{array}
        \right)    \left(
        \begin{array}{c}
         \rho^{++}\\
         \rho^{\prime++} \\
        \chi^{++}\\
        \chi^{\prime++}
      \end{array}
        \right).\crn
    \label{hdhdh01}\eea
Because the contributions from neutral Higgs mediations only relate
  to $\rho^{\prime 0}$,  so the contribution to the $e^{+}e^{-}\rightarrow H^{++}H^{--}$ amplitude
  is  proportional to
 $$ \frac{gm_e}{m_Wc_{\gamma}}\times m_W c_{\gamma}=\frac{m_e e^2}{s^2_{\theta_W}}. $$
This contribution is smaller than one
 from  neutral gauge boson mediation a
 factor of $m_e/\sqrt{s}$,  so we can neglect it.

The Higgs-Higgs-gauge boson vertices come from the covariant
kinetic terms of the Higgses
\bea \mathcal{L}^{\mathrm{kinetic}}_{\mathrm{H}}&=& \sum_{H}
\left(\mathcal{D}_{\mu}H\right)^{\dagger}\mathcal{D}^{\mu}H, \crn\rightarrow&&
\frac{ig}{2}\left(-\frac{2}{\sqrt{3}}V^{8\mu}+\frac{\sqrt{2}t}{\sqrt{3}}B^{\mu} \right)
\left(\rho^{--}\partial_{\mu}\rho^{++}+\rho^{\prime--}\partial_{\mu}\rho^{\prime++}\right)\crn
&-&\frac{ig}{2}\left(-V^{3\mu}+\frac{1}{\sqrt{3}}V^{8\mu}-\frac{\sqrt{2}t}{\sqrt{3}}B^{\mu} \right)
\left(\chi^{--}\partial_{\mu}\chi^{++}+\chi^{\prime--}\partial_{\mu}\chi^{\prime++}\right)+\mathrm{H.c}.
\nn\eea
 The interactions among neutral gauge bosons and DCHs can be written as
\bea
 \mathcal{L}_{HHV^0}&=& i 2e
A^{\mu}\left(\rho^{--}\partial_{\mu}\rho^{++}+\rho^{\prime--}\partial_{\mu}\rho^{\prime++}+
\chi^{--}\partial_{\mu}\chi^{++}+\chi^{\prime--}\partial_{\mu}\chi^{\prime++}\right)\crn &+&
\frac{ig}{2\sqrt{3}}\left[-\left(c_{\zeta}+\frac{(2t^2-3)s_{\zeta}}{\sqrt{2t^2+3}}\right)Z^{\mu} +
\left(s_{\zeta}-\frac{(2t^2-3)c_{\zeta}}{\sqrt{2t^2+3}}\right)Z'^{\mu}\right] \crn&\times&
\left(\rho^{--}\partial_{\mu}\rho^{++}+\rho^{\prime--}\partial_{\mu}\rho^{\prime++}\right)\crn &+&
\frac{ig}{2\sqrt{3}}\left[\left(c_{\zeta}-\frac{(2t^2-3)s_{\zeta}}{\sqrt{2t^2+3}}\right)Z^{\mu}
-\left(s_{\zeta}+\frac{(2t^2-3)c_{\zeta}}{\sqrt{2t^2+3}}\right)Z^{\prime\mu}\right] \crn&\times&
\left(\chi^{--}\partial_{\mu}\chi^{++}+\chi^{\prime--}\partial_{\mu}\chi^{\prime++}\right)+\mathrm{H.c.},
\label{Hkinetic}\eea
where $\mathcal{D}_{\mu}=\partial_{\mu}-i g V^a_{\mu}T^a -ig' X T^9B_{\mu}$, $T^a=\frac{1}{2}
\lambda^a$ or $-\frac{1}{2} \lambda^{a*}$ corresponding to triplet or anti-triplet representation
of Higgses, $T^9=\frac{1}{\sqrt{6}} \mathrm{diag}(1,1,1)$. In order to find the couplings of $Z, Z^\prime$ bosons with the DCHs,
we have to change the basis $(\rho^{--}, \rho^{\prime--},  \chi^{--},  \chi^{\prime--})$ into the
physical mass states $( G^{--},  H^{--}_1,  H^{--}_2,  H^{--}_3)$.
 Based on (\ref{massd1}),  if we ignore the suppressed
 terms containing a factor of $m^2_W/m^2_V$,   we can estimate the $H^{--}H^{++}V^0$  couplings.
In the limit $\Lambda_{11}\simeq 1$, $\Lambda_{12}=\Lambda_{13}\rightarrow 0$, the couplings of two different DCHs  with gauge bosons
are very suppressed.  So we only investigate the couplings of $H^{++}_iH^{--}_i V$. These  couplings
are almost independent to $\Lambda_{ij}$ or  masses of DCHs, as given in  Table \ref{HHVvertex}.
\begin{table}[h]
  \centering
  \begin{tabular}{|c|c|c|c|}
  \hline
  $V^{\mu}H^{--}_iH^{++}_i(p+p')_{\mu}$ & $A^{\mu}$ & $Z^{\mu}$ & $Z^{\prime\mu}$ \\
  \hline
  $H^{\pm\pm}_1$ &$2ie $  & $\frac{ig}{2\sqrt{3}} \left(\frac{2t^2-3}{\sqrt{2t^2+3}}s_{\zeta}-c_{\zeta}\right)$&
  $\frac{ig}{2\sqrt{3}} \left(\frac{2t^2-3}{\sqrt{2t^2+3}}c_{\zeta}+s_{\zeta}\right)$  \\
  \hline
$H^{\pm\pm}_{2,3}$ & $2ie$ & $\frac{ig}{2\sqrt{3}}
\left(\frac{2t^2-3}{\sqrt{2t^2+3}}s_{\zeta}+c_{\zeta}\right)$ &
$\frac{ig}{2\sqrt{3}} \left(\frac{2t^2-3}{\sqrt{2t^2+3}}c_{\zeta}-s_{\zeta}\right)$ \\
\hline
\end{tabular}
  \caption{Couplings of DCHs with neutral gauge bosons.}\label{HHVvertex}
\end{table}
\section{\label{signaldch} Signal of Doubly charged Higgses in $e^+e^-$ colliders}
\begin{figure}
  \centering
  \includegraphics[width=12cm]{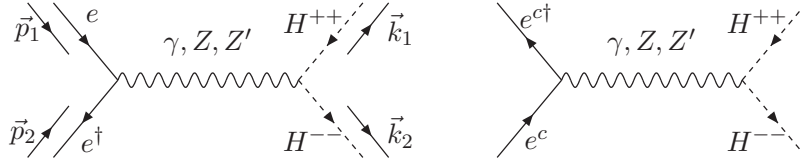}\\
  \caption{Feynman diagrams for production of
$H^{++}$ and its decays in $e^+e^-$ colliders.}
\label{dhiggsdcay1}
\end{figure}

 In a $e^{+}e^{-}$ collider, the reaction $e^{+}e^{-}\rightarrow
 H^{++}H^{--}$ may involve  the mediations of virtual neutral
 particles such as Higgses, gauge bosons. But main contributions
  relate to only  neutral gauge bosons, as the Feynman diagrams shown in  the figure \ref{dhiggsdcay1}.

In the center mass (CM) frame, the differential cross section for each DCH is given by
 \be \frac{d\sigma}{d(\cos\theta)}= \frac{1}{32\pi s} \sqrt{1-
 \frac{4m^2_{H^{\pm\pm}}}{s}}\left|\overline{\mathcal{M}} \right|^2,
\label{dcrossection1}\ee
 where $s=(p_{1}+p_{2})^2=E^2_{\mathrm{cm}}$ and $
 \mathcal{M}$ is the scattering amplitude, $\theta$ is the angle
 between $\vec{k}_1$ and $\vec{p}_1$. The detail calculation is
 shown in the appendix \ref{tdtx1}. The final result is
\bea \frac{d\sigma}{d(\cos\theta)}&=& -\frac{s}{32\pi} \sqrt{1-
 \frac{4m^2_{H^{\pm\pm}}}{s}}
\times \left(|\lambda_L|^2+|\lambda_R|^2\right)
\left(1+\cos^2\theta\right),\label{dcross2}\eea where
\bea \lambda^{H_1}_{L} &=& \sum_{a} \frac{G^a_LG^a_H}{s-m^{a2}_V+
i m^a_V \Gamma_a} \crn &=&e^2\times \left[\frac{2}{s}+
\frac{\left(c_{\zeta}+\frac{3s_{\zeta}}{\sqrt{2t^2+3}}\right)\left(\frac{2t^2-3}{\sqrt{2t^2+3}}s_{\zeta}-c_{\zeta}\right)}{12
s^2_{\theta_W}\left(s-m^2_Z+im_Z\Gamma_Z\right)}+
\frac{\left(-s_{\zeta}+\frac{3c_{\zeta}}{\sqrt{2t^2+3}}\right)
\left(\frac{2t^2-3}{\sqrt{2t^2+3}}c_{\zeta}+s_{\zeta}\right)}{12
s^2_{\theta_W}\left(s-m^2_{Z'}+im_{Z'}\Gamma_{Z'}\right)}\right], \crn
 \label{lamdal1}\eea
where $a=\gamma,~Z, ~Z'$ , total width of the $Z'$  is given
 in the appendix \ref{apllv} and
 \bea
   \lambda^{H_1}_{R} &=&\sum_{a}
\frac{G^a_RG^a_H}{s-m^{a2}_V+i m^a_V \Gamma_a} \crn &=&e^2 \left[\frac{2}{s}-
\frac{\left(\frac{2t^2-3}{\sqrt{2t^2+3}}s_{\zeta}-c_{\zeta}\right)
\left(-\frac{3s_{\zeta}}{\sqrt{2t^2+3}}+c_{\zeta}\right)}{12
s^2_{\theta_W}\left(s-m^2_Z+im_Z\Gamma_Z\right)}-
\frac{\left(\frac{2t^2-3}{\sqrt{2t^2+3}}c_{\zeta}+s_{\zeta}\right)
\left(\frac{3c_{\zeta}}{\sqrt{2t^2+3}}+s_{\zeta}\right)}{12
s^2_{\theta_W}\left(s-m^2_{Z'}+im_{Z'}\Gamma_{Z'}\right)}\right].\crn
\label{lamdar1}\eea
Here $G^a_L, G^a_H$ are the couplings of the neutral gauge bosons with two leptons and two DCHs,
respectively.

Similarly,   in the case of $H^{\pm\pm}_{2,3}$  we have \bea \lambda^{H_{2,3}}_{L} =e^2\times \left[\frac{2}{s}+
\frac{\left(c_{\zeta}+\frac{3s_{\zeta}}{\sqrt{2t^2+3}}\right)
\left(\frac{2t^2-3}{\sqrt{2t^2+3}}s_{\zeta}+c_{\zeta}\right)}{12
s^2_{\theta_W}\left(s-m^2_Z+im_Z\Gamma_Z\right)}+
\frac{\left(\frac{3c_{\zeta}}{\sqrt{2t^2+3}}-s_{\zeta}\right)
\left(\frac{2t^2-3}{\sqrt{2t^2+3}}c_{\zeta}-s_{\zeta}\right)}{12
s^2_{\theta_W}\left(s-m^2_{Z'}+im_{Z'}\Gamma_{Z'}\right)}\right]~~
 \label{lamdal2}\eea
 and
 \bea
   \lambda^{H_{2,3}}_{R} =e^2
\left[\frac{2}{s}-
\frac{\left(\frac{2t^2-3}{\sqrt{2t^2+3}}s_{\zeta}+c_{\zeta}\right)
\left(c_{\zeta}-\frac{3s_{\zeta}}{\sqrt{2t^2+3}}\right)}{12
s^2_{\theta_W}\left(s-m^2_Z+im_Z\Gamma_Z\right)}-
\frac{\left(\frac{2t^2-3}{\sqrt{2t^2+3}}c_{\zeta}-s_{\zeta}\right)
\left(\frac{3c_{\zeta}}{\sqrt{2t^2+3}}+s_{\zeta}\right)}{12
s^2_{\theta_W}\left(s-m^2_{Z'}+im_{Z'}\Gamma_{Z'}\right)}\right].\label{lamdar2}\eea
The total cross section  is
 \be \sigma= \frac{s}{12\pi} \sqrt{1-
 \frac{4m^2_{H^{\pm\pm}}}{s}}
\times \left(|\lambda_L|^2+|\lambda_R|^2\right).
\label{tcrossec1}\ee
The above process happens only when $\sqrt{s}> 2
 m_{H^{\pm\pm}}>400$ GeV from the prediction of the SUSYRM331.
 \begin{figure}[h]
  \centering
\begin{tabular}{cc}
\epsfig{file=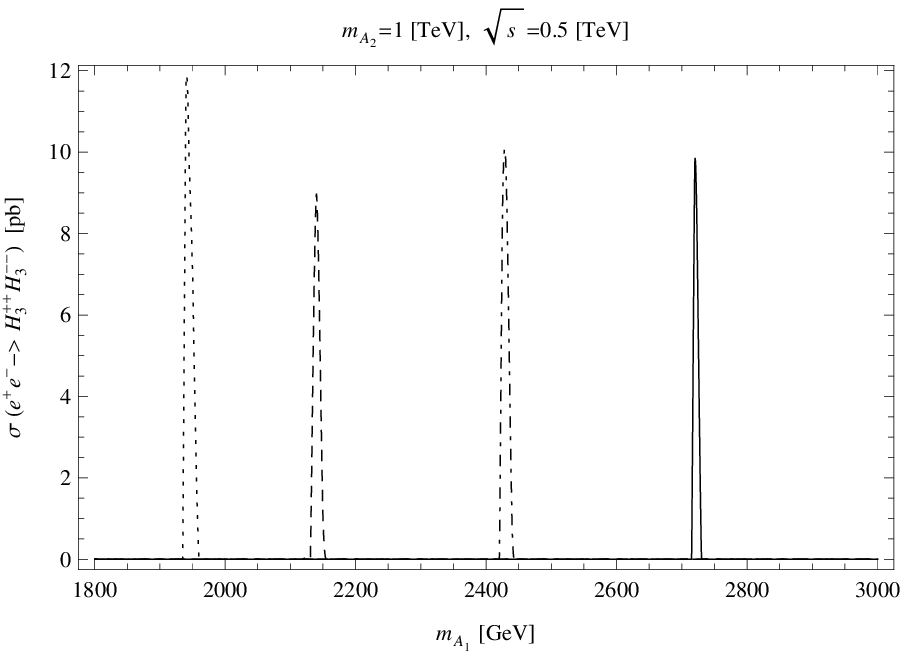,width=0.4\linewidth,clip=}
&
\epsfig{file=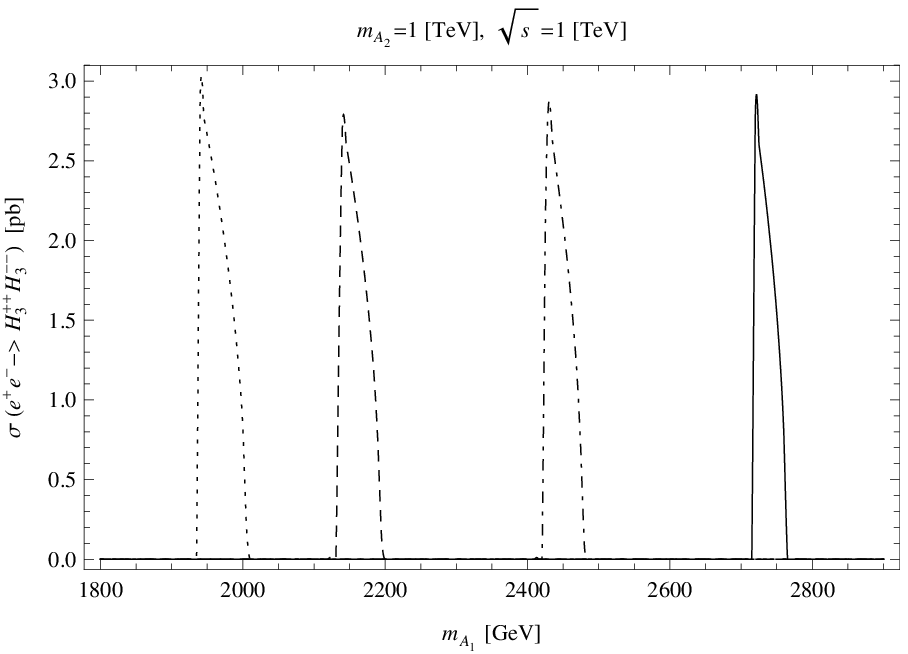,width=0.4\linewidth,clip=}
\\
\epsfig{file=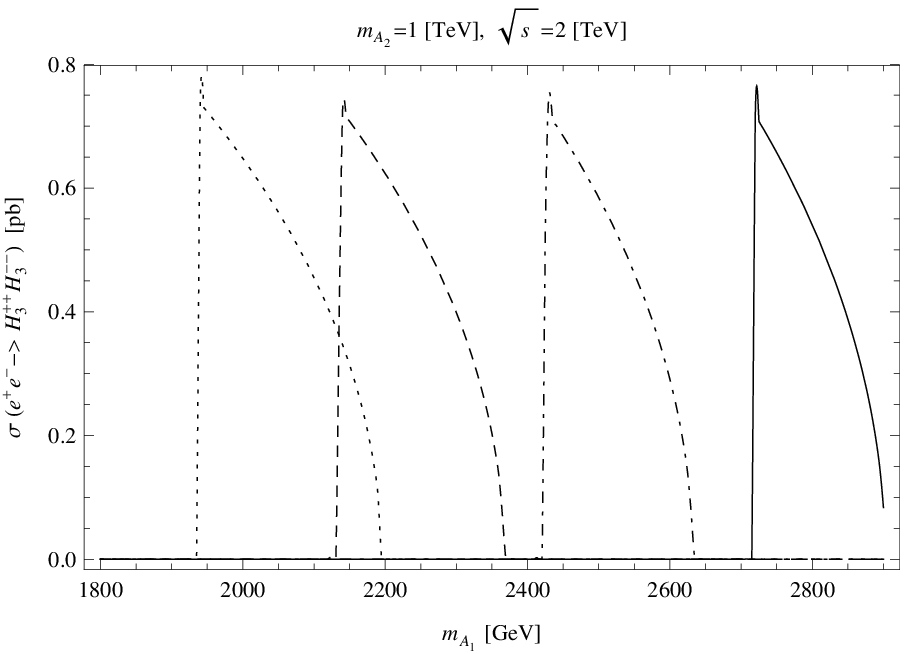,width=0.4\linewidth,clip=}
&
\epsfig{file=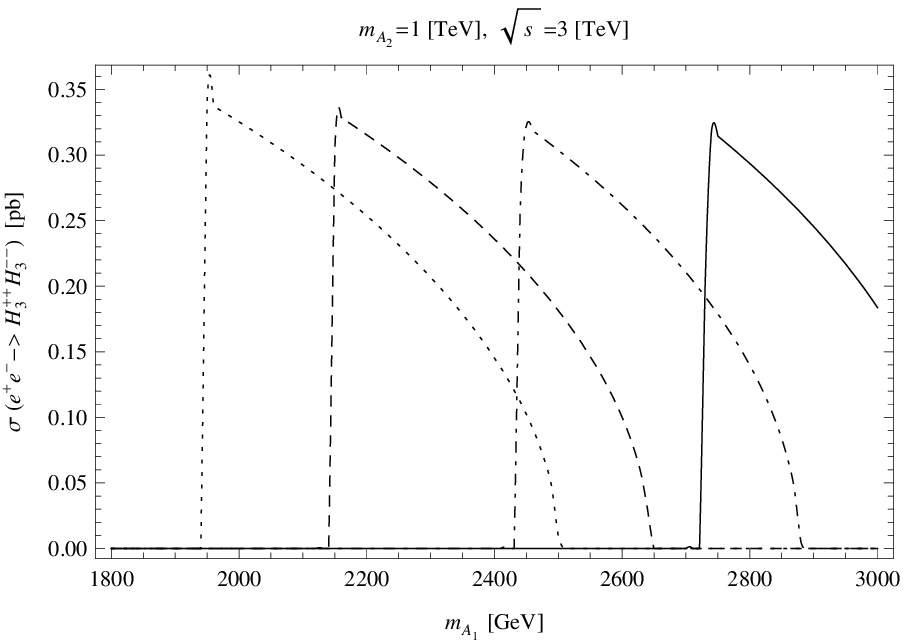,width=0.4\linewidth,clip=}
\\
\end{tabular}
  \caption{Plots of the production cross sections of the lightest DCH $H^{\pm\pm}_{3}$   as functions of
  $m_{A_1}$ with different colliding energies.  The values of $m_V$ are
    $m_V =2,~2.2,~2.5$ and  $2.8$ TeV represented by dotted, dot-dashed,  dashed and black curves, respectively.}
   \label{sigmaHcc3}
\end{figure}

To determine the signals of the lightest DCH, we firstly investigate the
dependence of the cross section of the process $e^+e^-\rightarrow
H^{++}_3H^{--}_3$ on the fixed collision energies of 0.5, 1, 2 and 3 TeV, as shown in the figure \ref{sigmaHcc3}. For $\sqrt{s}=0.5$ TeV, with each fixed values of $m_V$ there
exists a very small range of $m_{A_1}$ corresponding to the
creation of $H^{\pm\pm}_3$. This is because of the fact that the
small $m_{A_1}$ will create the tachyon DCH while the
large values will make   masses of the DCHs be larger  than the allowed kinetic condition. The cross section in this case
can reach few pbs.   For larger
$\sqrt{s}$, the cross sections decrease but still reach $\mathcal{O}$(0.1) pb.  One of the most important property of the lightest SUSYRM331 DCH is that its mass characterizes the difference between two parameters $m_{A_1}$ and $m_V$. Hence the signal of DCH requires the nearly degeneration between these two mass, $|m_{A_1}-m_V|<100$ GeV when  $\sqrt{s}\leq 1$ TeV. We also can see that the low colliding enegies give a rather large cross section for creating the lightest DCH.
 \begin{figure}[h]
  \centering
\begin{tabular}{cc}
\epsfig{file=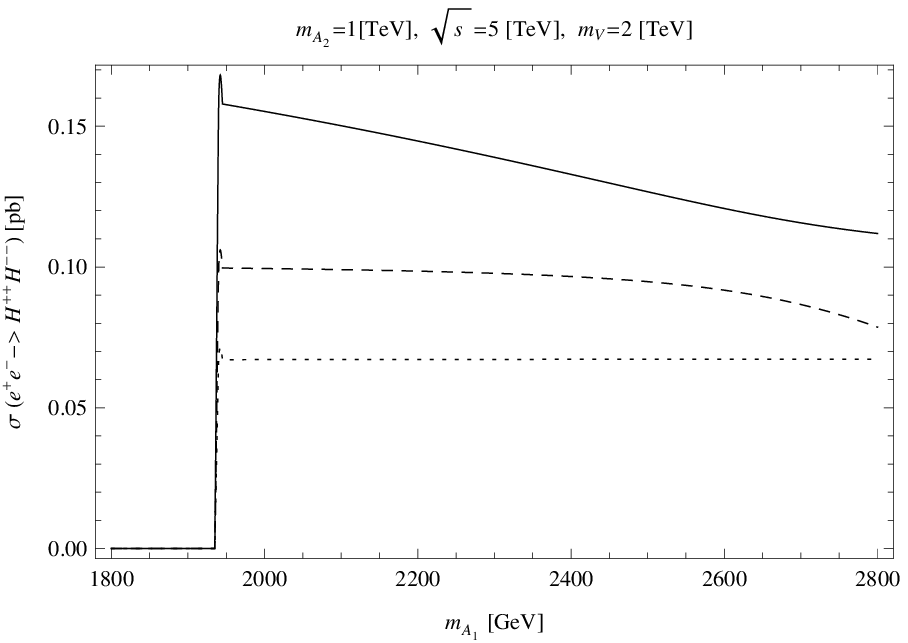,width=0.45\linewidth,clip=}
&
\epsfig{file=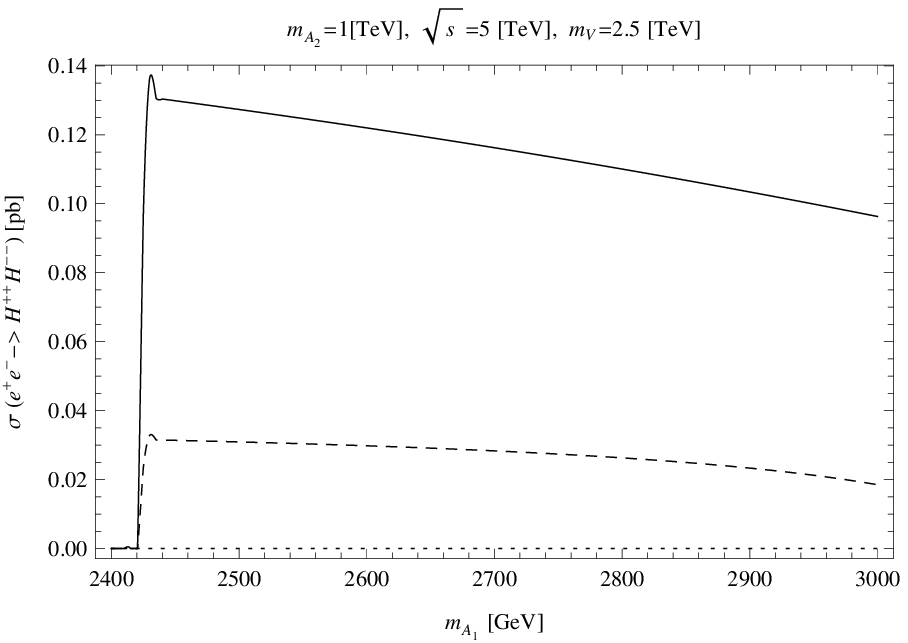,width=0.45\linewidth,clip=}
\\
\end{tabular}
  \caption{Total cross sections of creating three DCHs in $e^{+}e^{-}$ colliders as functions of $m_{A_1}$ in very high colliding energy of 5 TeV and   $m_{V}= 2$ (2.5) TeV. We denote: the heaviest-dotted curves,
    the second heavy-dashed curves and the lightest-black curves.}
   \label{sigma3Hcc}
\end{figure}

The heavier DCHs may be
created with very high collision energies, i.e higher than 4 TeV. For illustration,  the figure  \ref{sigma3Hcc} represents the
total cross sections $\sigma (e^{+}e^{-}\rightarrow H^{++}H^{--})$
of three DCHs  in the CM energy of $\sqrt{s}=5$ TeV although it goes beyond the maximal energy that both ILC and CLIC can reach.  Because all $m_{A_1},m_{A_2},m_V \gg m_{W}$, the
cross sections of  the DCHs depend weakly on the change of $m_{A_1}$. Apart from the $m_{H^{\pm\pm}_{2,3}}$, the $m_{A_1}$ affects only the decay width of the $m_{Z'}$ which gives small contribution to the cross section in the limit of very large SUSY and $SU(3)_L$ scales. The value of  $m_V=2$ TeV gives $m_{H^{\pm\pm}_1}\simeq \sqrt{m^2_{A_2}+m^2_{V}}=\sqrt{s}$, leading to a rather small of cross section of $\mathcal{O}(10^{-2})$ pb (the dotted curve in the left-panel) and does not depend on $m_{A_1}$.  While the value of $m_V=2.5$ TeV gives $ m_{H^{\pm\pm}_1}>\sqrt{s}$, the  $H^{\pm\pm}_1$ cannot appear.  For $H^{\pm\pm}_2$, as explained above,  its mass is also independent with small $m_{A_1}$. Furthermore,  all couplings and gauge boson masses relating with the cross
sections are independent with  the mentioned range of  $m_{A_1}$. So  the $\lambda^{H}_{L,R}$ shown
 in (\ref{lamdal1})-(\ref{lamdar2}) will become constant too, giving the same property of the  cross section of this DCH. But it is very sensitive with $m_V$.  In particular, it can get value of 0.1 pb with $m_V=2$ TeV but reduces to $0.03$ pb $m_V=2.5$ TeV.  When $\sqrt{s}=5$ TeV, the cross section of the lightest DCH is 0.1 pb for all masses satisfying the kinetic condition, rather smaller than other cases of $\sqrt{s}<3$TeV.
 \begin{figure}[h]
  \centering
\begin{tabular}{cc}
\epsfig{file=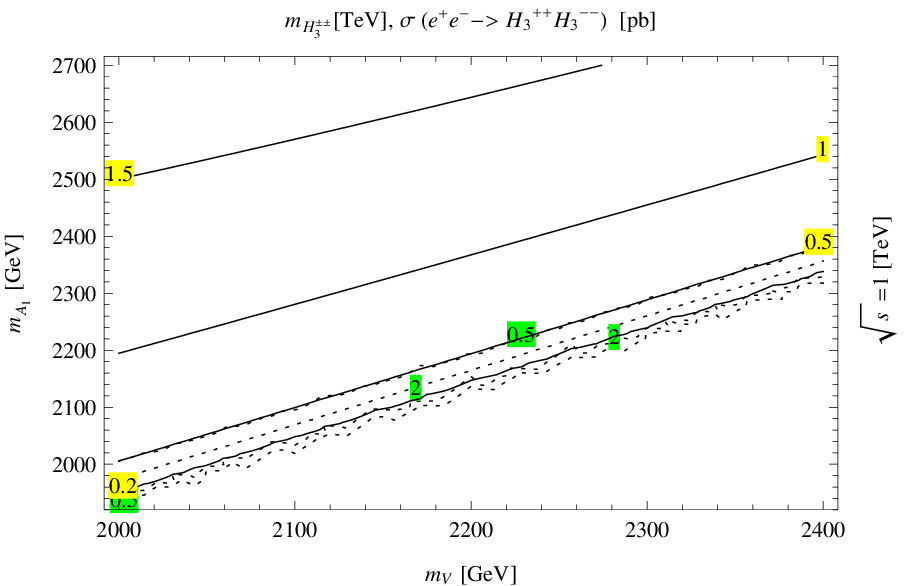,width=0.5\linewidth,clip=}
&
\epsfig{file=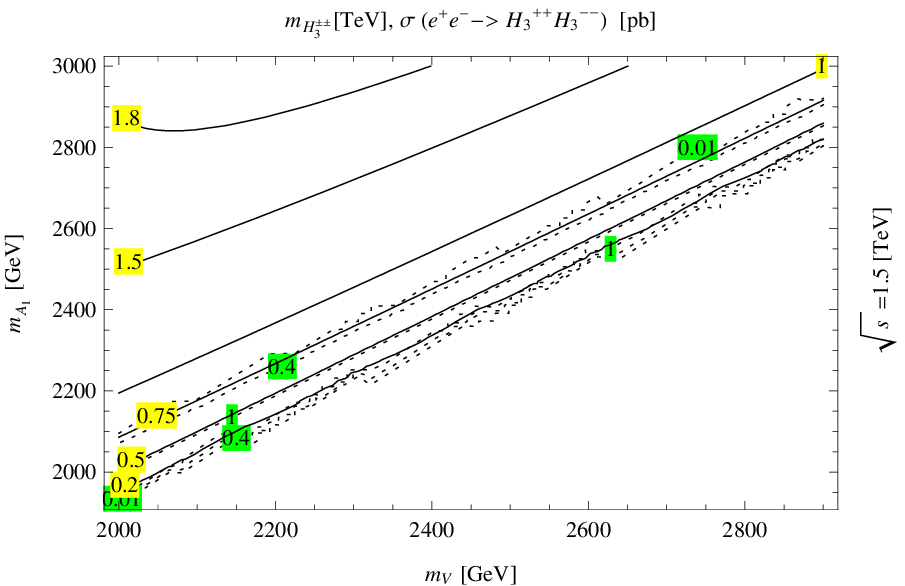,width=0.5\linewidth,clip=}
\\
\epsfig{file=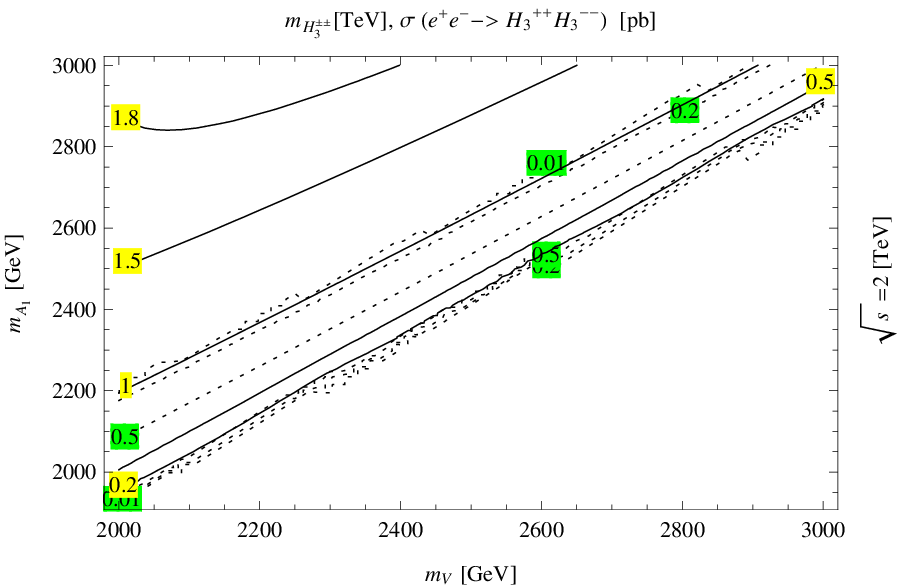,width=0.5\linewidth,clip=}
&
\epsfig{file=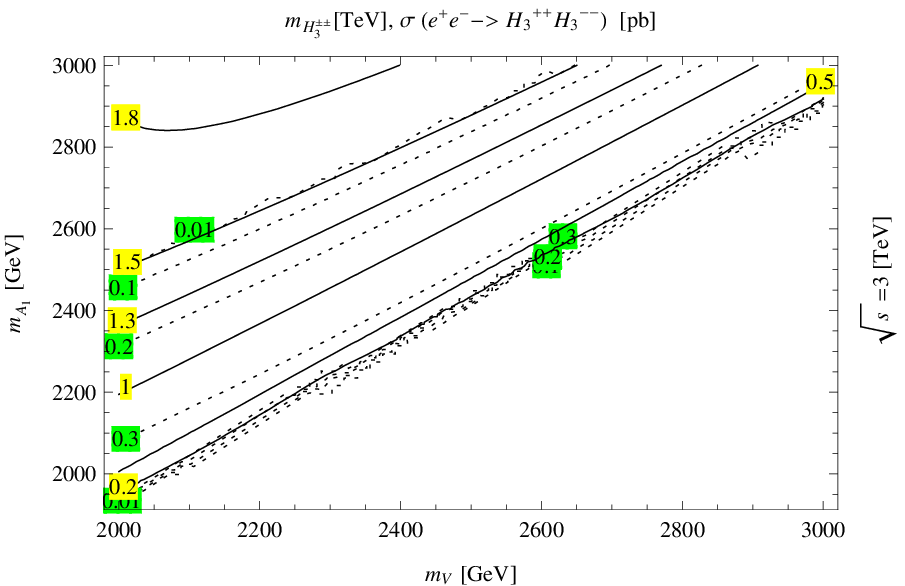,width=0.5\linewidth,clip=}
\\
\end{tabular}
  \caption{ Contour plots of the mass and the production cross section of  the lightest DCH  in  $e^{+}e^{-}$ colliders  as functions of $m_{V}$ and  $m_{A_1}$ with different colliding energies of $1,1.5,2$ and 3 TeV. The mass and the cross section are represented by  black
    and dotted  curves, respectively.}
   \label{sigmavsmass}
\end{figure}

The two figures \ref{sigmaHcc3} and \ref{sigma3Hcc} only help us see
 the maximal
  values of the cross sections for creation the DCHs. The above discussion does not
   include the lower bound on  the lightest DCH mass. The figure  \ref{sigmavsmass}
  is used to  estimate values of the cross sections $\sqrt{s}=1-3$ TeV, including
   both condition of lower DCH mass bound and the allowed kinetic condition for
   creating heavy physical DCHs. All panels in both figures have same
interesting properties. When a DCH mass
approaches the limit of the kinetic allowed value, the
corresponding  cross section will
decrease  to zero.  This explains why the
contours of these two quantities almost overlap each other in the
limits of $m_{H^{\pm\pm}}\rightarrow \sqrt{s}/2$ and cross section
$\sigma\rightarrow 0$.

 \begin{figure}[h]
  \centering
\begin{tabular}{cc}
\epsfig{file=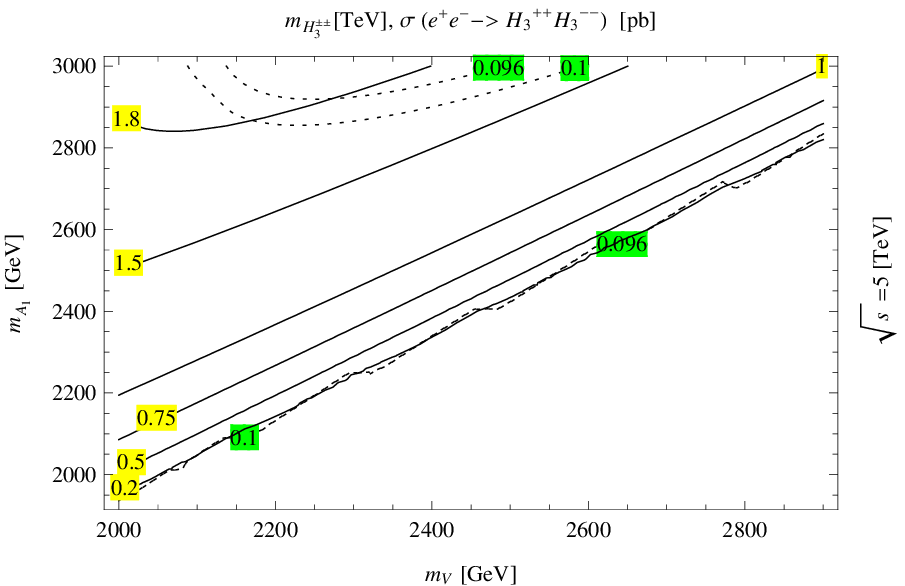,width=0.5\linewidth,clip=}
&
\epsfig{file=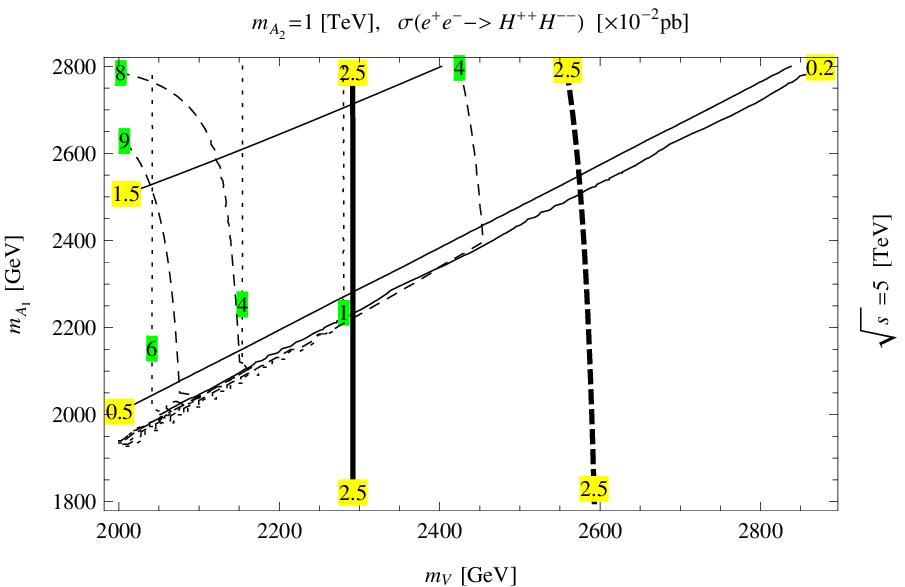,width=0.45\linewidth,clip=}
\\
\end{tabular}
  \caption{ Contour plots of  production cross sections of  DCHs
  in $e^{+}e^{-}$ collider as functions of $m_{V}$ and  $m_{A_1}$ at colliding energy of 5 TeV. The left panel  focuses  on the lightest, where    the dotted and black curves describe  the cross section and  mass, respectively.
      The right panel represents the cross sections
       of the  remain two DCHs, in particular the second heavy and the heaviest are
       described by dashed  and  dotted curves, respectively. In addition, dashed and
       black thick curves  represent the maximal mass values of the DCHs allowed by the kinetic condition.}
   \label{csigma3hcclr}
\end{figure}

In the figure \ref{sigmavsmass}, the cross section can reach  the value
of few pbs with $\sqrt{s}=1$ TeV if the lower bound of the DHC mass of $200$ GeV is considered. In general the value of few pb can be reached for searching the very light DCH with mass below 250 GeV  in  colliding energies in range of 0.5-1 TeV. These values of the cross section are much smaller than the maximal value for $\sqrt{s}=0.5$ TeV shown in the figure \ref{sigmaHcc3}.
And  the region of the parameter space
allowed for the DCH appearance is very narrow, implying the
degeneration of $m_{A_1}$ and $m_V$. With $\sqrt{s}=1.5-2$
TeV, the lightest DCH with mass $500$ GeV $<m_{H^{\pm\pm}_3}<750$
GeV may be detected with the corresponding $ \sigma > 0.4$ pb. More interesting, it can reach
1 pb, twice larger than the case of $\sqrt{s}=1$ Tev, if the mass is  around 500 GeV. For $\sqrt{s}=3.0$, The cross sections for $m_{H^{\pm\pm}_3} > 0.5$ TeV  is not larger than 0.3 pb.
 This value is close to the maximal value shown in the figure \ref{sigmaHcc3}. From this we can conlude that the largest cross section for searching the lightest DCH with mass from $0.5$ TeV to 0.75 TeV corresponds to the intermediate values of $\sqrt{s}$ from 1.5 TeV to 2 TeV.

The figure  \ref{csigma3hcclr} shows the rather small  cross sections of creating all DCHs, when $\sqrt{s}=5$ TeV. For the lightest, the maximal is below 0.1 pb,  while the two others the values is order of $10^{-2}$ pb.

All above numerical investigations show that the production cross sections  of the lightest DCHs in $e^+e^-$ colliders can be reach values of
$10^{-1}-$ few pbs, depending on the DCH mass and the collision energy. This will be  a good signal for the detection the lightest DCH in near future colliders
\cite{ILC1,ILC2,CLIC1,CLIC2}. In particularly, the ILC with the collision energy of 0.5-1 TeV
corresponding to the integrated luminosity of $500-1000~ \mathrm{fb^{-1}}$ \cite{ILC1,ILC2}, the number of events
for creation the lightest DCH will be around $5\times 10^5-10^6$ corresponding to  the DCH mass range of 200-500 GeV. With the CLIC \cite{CLIC1,CLIC2} where the
collision energy will increase up to 3 TeV or more, the  lightest DCH may be observed with larger range of mass. Furthermore,  the
estimated integrated luminosity targets will be  1.5 $\mathrm{ab}^{-1}$ at 1.4
 (1.5) TeV and 2 $\mathrm{ab}^{-1}$ at 3 TeV collision energy.
 The DCH with mass below 750 GeV gives the best signal with $\sqrt{s}=1.5-2$ TeV, where the observed event numbers can reach   $6\times 10^5-1.5\times 10^6$.  With $s=3$ TeV  the  maximal number of
 events reduces to  $6\times 10^5$.  When the collision energy is high enough to create heavy DCH, the event numbers reduce to $10^4$ corresponding to the luminosity of 1 $\mathrm{ab^{-1}}$.

\section{\label{con} Conclusions}
We have investigated the Higgs sector of the SUSYRM331 model where
the DCHs are specially concentrated on as one of signals to look
for new physics at $e^+e^-$  colliders. Here,  masses of neutral
CP-even Higgses,  DCHs and the cross section of creating DCHs at $e^+e^-$
colliders can be represented according to five unknown parameters:
two masses $m_{A_1,A_2}$ of neutral CP-odd Higgses characterizing the soft scale; mass of singly
heavy gauge boson $m_V$-the $SU(3)_L$ breaking scale; both ${\gamma}$ and ${\beta}$ relating to the ratios of
the VEVs.  This choice of parameters helps us more easily to discuss on the relations among not only  particle
masses but also the breaking scales of the model. We have found the exact
 condition
$ \frac{\left(m^2_{A_1}+m^2_W\right) c_{2\gamma}}{m^2_V}
 <c_{2\beta}<\frac{m^2_W  c_{2\gamma}}{m^2_V+m^2_{A_2}}<0$,
  which must be satisfied to avoid tachyons of  the DCHs at the tree level. The
 numerical investigation of  the DCHs as functions of $m_{A_1}$ and $m_V$
  shows that even with very large values of  all
  $m_{A_1},~ m_{A_2}$ and $m_V$  there may still exist a light DCH if value $m_{A_1}$
   is enough close to that of $m_V$,  being consistent
  with the relation $ 0<-c_{2\beta}\left[c_{2\beta} m^2_V-
c_{2\gamma} (m^2_W +m^2_{A_1})\right] \sim \mathcal{O}(m^2_W)$ found
by our analysis. The constraint on the decay $H^0_1\rightarrow \gamma\gamma$
gives the lower bound on the mass of the DCH of about 200 GeV, the same as the experimental value given by CMS.
Finally, we have investigated
the possibility of creating  the DCHs in $e^+e^-$ colliders with collision energies from 1 to 3 TeV, then
 indicated that
only the lightest DCH may be created. The production cross sections are from 0.1 pb to few pbs,  depending on the mass range and the collison energy.
Because the SUSYRM331 is valid in the limit of very large $SU(3)_L$ scale, two other  DCHs always have masses above 2 TeV, therefore they do not appear unless the collision energies  are higher than 4 TeV.
  Anyway, they will give small cross sections for all three DCHs, with order of  $\mathcal{O}(10^{-2})$ pbs for the two heavier DCHs and  0.1 pb for the lightest. The two heavy DCHs are difficult to observed in  the LHC, ILC and CLIC.
On the other hand,  the lightest DCH, which only decays to same-sign pair of
charged tauons,  will  give the most promising  signal for searching
it in $e^+e^-$  colliders such as  the ILC and CLIC.
If  it is  detected,  the numerical investigation  in this work will give interesting
 informations of parameters such as $m_{A_1}$ and $m_V$.
\section*{Acknowledgments}
This research is funded by Vietnam  National
Foundation for Science and Technology Development (NAFOSTED) under
grant number 103.01-2014.69.
\appendix
\section{\label{tdtx1} Cross section of $e^{+}e^{-}\rightarrow H^{++}H^{--}$}
\begin{figure}
  \centering
  \includegraphics[width=12cm]{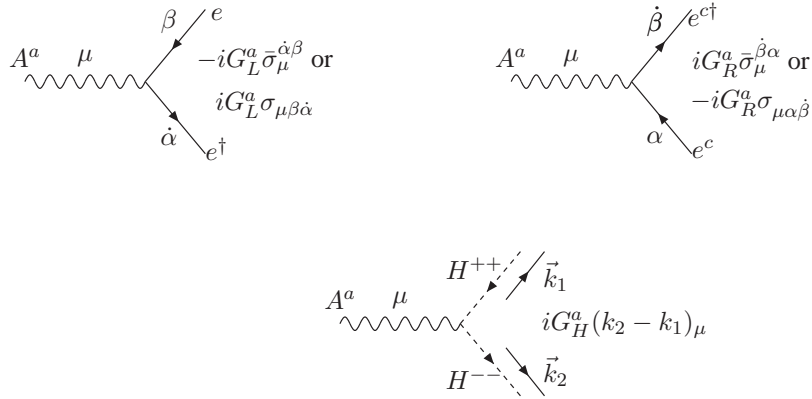}\\
  \caption{Feyman rules for interacting vertices of
$llA^a$ and $H^{++}H^{--}A^a$ where $A^a$ is a physical neutral
gauge boson.} \label{Feymanrule1}
\end{figure}

Lagrangian for process  $e^{+}e^{-} \rightarrow H^{++}H^{--}$ can
be written in terms of two-component spinor
 \bea \mathcal{L}_{eeV^0}=
-A^{a}_{\mu}\left(G^a_L e^{\dagger}\bar{\sigma}^{\mu}e-G^a_R
e^{c\dagger}\bar{\sigma}^{\mu}e^c\right)+
\sum_{H=H^{\pm\pm}_{1,2,3}}G^a_H A^a_{\mu} \left(
H^{++}\partial_{\mu}H^{--}-\mathrm{H.c.}\right), \label{eevin}\eea
where $A^a_{\mu}=A_{\mu},~Z_{\mu},~Z'_{\mu}$, $G^{a}_L$ and
$G^a_R$ are given in table \ref{eeVvertices1}, $G^a_H$  are given
in table \ref{HHVvertex}. Feynman rules  can be found in
\cite{martin1} for example, the detail shown in the figure
\ref{Feymanrule1}.
Denoting $k=k_2-k_1$, then the total amplitude for each
DCH is written as
\bea i\mathcal{M}_H&=&-i
\left(x^{\dagger}_2[\bar{\sigma}.k]x_1\right)\sum_{a}
\frac{G^a_LG^a_H}{(p_1+p_2)^2-m^{a2}_V}-i
\left(y_2[\sigma.k]y^{\dagger}_1\right)\sum_{a}
\frac{G^a_LG^a_H}{(p_1+p_2)^2-m^{a2}_V} \crn
&\equiv& -i
\left(x^{\dagger}_2[\bar{\sigma}.k]x_1\right)\lambda_L-i
\left(y_2[\sigma.k]y^{\dagger}_1\right)\lambda_R, \label{bd1}\eea
where $m^a_V=0,m_Z,m_{Z'}$ corresponding to photon, $Z$, $Z'$
bosons.  Squaring the amplitude and summing over the electron and
positron spins, we have
\bea \left|\overline{\mathcal{M}}
\right|^2&=&\sum_{s_1,s_2}\mathcal{M}_H^{\dagger}\mathcal{M}_H
\crn&=& 2\left(|\lambda_L|^2+|\lambda_R|^2\right)\left[
2(p_1.k)(p_2.k)-(p_1.p_2) k^2\right]+ 2 \Re(\lambda_L^* \lambda_R)
m_e^2 k^2.\label{squareamp1}\eea
Now we use the fact that $p_1^2=p_2^2=m^2_e\simeq 0$,
$k^2_1=k^2_2=m^2_{H^{\pm\pm}}$. Furthermore, all terms in
(\ref{squareamp1}) are invariant  under the Lorentz
transformation, so the result is unchanged when we use any
particular frame. Here we use the center mass frame where the
momenta of two initial particles are
\be p_{1\mu}=\left( E,0,0,E\right), \hs p_{2\mu}=\left(E,0,0,-E\right)\label{cmframe}\ee with  $E=E_{cm}/2=\sqrt{s}/2$.
We define two pour-momenta of final particles as
$k_{1\mu}=(E_1,\vec{k}_1)$ and $k_{2\mu}=(E_2,\vec{k}_2)$. Using
the condition of four-momentum conversation, it is easy to prove
some following results
 \bea k^2&=&(k_1-k_2)^2=4m^2_{H^{\pm\pm}}-s,\hs
 p_1.k=-p_2.k=\frac{s\cos\theta}{2}\sqrt{1-\frac{4m^2_{H^{\pm\pm}}}{s}}
 \crn
 p_1.p_2&=& \frac{s}{2}-m^2_e\simeq \frac{s}{2}.\label{somere}\eea
Inserting all results (\ref{somere}) into (\ref{squareamp1}), we
obtain
\bea \left|\overline{\mathcal{M}} \right|^2&=&
-\left(|\lambda_L|^2+|\lambda_R|^2\right) s^2
\left(1+\cos^2\theta\right)+ 2 \Re(\lambda_L^* \lambda_R) m_e^2
\left(s-4m^2_{H^{\pm\pm}}\right)\crn
&\simeq&-\left(|\lambda_L|^2+|\lambda_R|^2\right) s^2
\left(1+\cos^2\theta\right).\label{squareamp2}\eea

\section{ Total width of the $Z'$ gauge boson}
For any particles $\phi$ (fermion,  gauge boson, scalar) in the model, we  define the corresponding covariant derivative
relating with neutral gauge bosons as
 $D_{\mu} \phi\equiv\left( \partial_{\mu}-i q_{\phi} A_{\mu} -i g^{\phi}_{Z} Z_{\mu}-ig^{\phi}_{Z'} Z'_{\mu}\right)\phi$.
 The analytic forms of  $g^{\phi}_{Z'}$ depend on the particular representation of $\phi$.
 Specially, we have
\begin{itemize}
\item  $\mathrm{SU(3)}_L$ singlet $$g^{\phi}_{Z'}=\frac{-gY_{\phi}c_{\zeta}t^2}{\sqrt{3(2t^2+3)}}  $$
  \item $\mathrm{SU(3)}_L$ triplet
  $$g^{\phi}_{Z'} =\frac{g}{2\sqrt{3}}\left(
                            \begin{array}{ccc}
                            - 2\left( s_{\zeta}+\frac{Y_{\phi}t^2c_{\zeta}}{\sqrt{2t^2+3}}\right) &0 & 0 \\
                              0 &s_{\zeta}-\frac{c_{\zeta}\left(2Y_{\phi}t^2+3\right)}{\sqrt{2t^2+3}}  & 0 \\
                              0 & 0 &s_{\zeta}+\frac{c_{\zeta}\left(-2Y_{\phi}t^2+3\right)}{\sqrt{2t^2+3}} \\
                            \end{array}
                          \right)
  $$
  \item  $\mathrm{SU(3)}_L$  anti-triplet  $$g^{\phi}_{Z'}=\frac{g}{2\sqrt{3}}\left(
                            \begin{array}{ccc}
                             2\left( s_{\zeta}-\frac{Y_{\phi}t^2c_{\zeta}}{\sqrt{2t^2+3}}\right) &0 & 0 \\
                              0 &-s_{\zeta}-\frac{c_{\zeta}\left(2Y_{\phi}t^2-3\right)}{\sqrt{2t^2+3}}  & 0 \\
                              0 & 0 &-s_{\zeta}-\frac{c_{\zeta}\left(2Y_{\phi}t^2+3\right)}{\sqrt{2t^2+3}} \\
                            \end{array}
                          \right)  $$
   \item  $\mathrm{SU(3)}_L$  adjoint representation relates with gauge bosons and their superpartners
   only.  The standard  couplings of three gauge bosons can be written as
   $i g^{V}_{Z'} \left[ g^{\mu\nu}(p-k_1)^{\sigma}\right.$ $\left.+g^{\sigma\nu}(k_1-k_2)^{\mu}+g^{\mu\sigma}(k_2-p)^{\nu}\right]$,
    where  $g^{VV'}_{Z'}$ shown in the table \ref{gzpvvp}. With gauginos, the vertices can also written in the form of
     $ i g^{\tilde{V}}_{Z'}Z'_{\mu}\tilde{V}^{\dagger}\bar{\sigma}^{\mu}\tilde{V}$, where
     $g^{\tilde{V}}_{Z'}=g^{V}_{Z'}$.
\begin{table}[h]
  \centering
  \begin{tabular}{|c|c|c|c|}
    \hline
    Gauge boson& $Z'W^+W^-$  &$Z'U^{++}U^{--}$ & $Z'V^+V^-$  \\
     \hline
     $g^{V}_{Z'}$, $(g^{\widetilde{V}}_{Z'})$& $\frac{g\sqrt{3}}{2}\left(-s_{\zeta}+
     \frac{c_{\zeta}}{\sqrt{2t^2+3}}\right)$ & $ -\frac{g\sqrt{3}c_{\zeta}}{\sqrt{2t^2+3}}$ & -$\frac{g\sqrt{3}}{2}
     \left(s_{\zeta}+\frac{c_{\zeta}}{\sqrt{2t^2+3}}\right)$  \\
    \hline
  \end{tabular}
  \caption{ Couplings of $Z'$ gauge boson with two gauge bosons (two gauginos) }\label{gzpvvp}
\end{table}
\end{itemize}
  Below we will calculate the partial decay width of $Z'$ into three different
  class of particles.  Analytic formulas can be found in \cite{manuel}.  For
   the purpose of estimation the total width decays of $Z'$ as simple as possible,
    we only consider the largest contribution to each class of particles. In addition,
    all particles as gauginos, sleptons, squarks receiving masses from  the soft terms are very heavy  so that
    $Z'$ cannot decay into.  We assume the similar situations for cases of exotic quarks.  The decay of the $Z'$
     relating with these particles deserve a further detailed study.  Here numerical values are used  :
     $s_{\zeta}= 0.155$, $ c_{\zeta}=0.988$  corresponding to the definition (\ref{dzeta}) in the case $m_V\gg m_{W}$.
     The value of $t$ follows the definition (\ref{tvalues}) with $s^2_{W}=0.231$.
 \subsection{Decay of $Z'$ to fermions pairs}
This kind of decay involves with the below Lagrangian \be
\mathcal{L}_{Z'ff}= \sum_{f}Z^{\prime\mu}\left(g^{f}_{Z'}
f^{\dagger}\overline{\sigma}_{\mu}f+ g^{f^c}_{Z'}
f^{c\dagger}\overline{\sigma}_{\mu}f ^c\right)=
\sum_{f}Z^{\prime\mu}\left( g^{f}_{Z'}
f^{\dagger}\overline{\sigma}_{\mu}f- g^{f^c}_{Z'}
f^{c}\sigma_{\mu}f ^{c\dagger}\right), \label{Zpfermion}\ee
where sum is over all fermions in the model that couple with $Z'$
and satisfy the kinetic condition $m_{Z'}> 2 m_{f}$. Formulas of
$g^{f}_{Z'}$ and $g^{f^c}_{Z'}$ were shown in the table
\ref{eeVvertices1}. The partial decay width corresponding to each
fermion is \cite{martin1}, \be \Gamma(Z'\rightarrow ff^+)=
\frac{N^f_c m_{Z'}}{24\pi}\left(
1-\frac{4m^2_f}{m^2_{Z'}}\right)^{1/2}
 \left[  \left( |g^f_{Z'}|^2+ |g^{f^c}_{Z'}|^2\right)\left(1-\frac{m^2_f}{m^2_{Z'}}\right)-6g^{f}_{Z'}g^{f^c}_{Z'}
 \frac{m^2_f}{m^2_{Z'}}\right] ,\label{Zpfermiondecay} \ee
 where $N^f_c$ is the  color factor being equal to 3 for quarks and 1 for all other
 fermions (leptons, quarks, Higgsinos and gauginos).
 \subsection{Decay of $Z'$ to scalar pairs}
 Lagrangian relating with these decay is
 \be \mathcal{L}_{Z'S_iS_i}=\sum_{S_i,S_j}ig^{S_{ij}}_{Z'} Z'_{\mu}\left[ S_i^{\dagger}
 \partial^{\mu}S_j- (\partial^{\mu}S_i^{\dagger})S_i\right], \label{zp2scalar}\ee 
 where $S_i$ stands for any scalars in the model . The Feynman rule is the same as the
 DCH shown in the figure \ref{Feymanrule1}, where $G^a_H\rightarrow i
 g^{S_{ij}}_{Z'}$. Non-zero values of $ g^{S_{ij}}_{Z'}$ for Higgses in the
 model are shown in the figure \ref{Zhh}.
\begin{table}
  \centering
\begin{tabular}{|c|c|c|c|}
  \hline
  Vertex & factor & Vertex & factor \\
  \hline
  $Z'H_1^{+}H_1^{-}$ & $\frac{-ig}{\sqrt{3}}\left(s_{\zeta}+\frac{t^2c_{\zeta}}{\sqrt{2t^2+3}}\right)$ &
   $Z'H_2^{+}H_2^{-}$ &  $\frac{-ig}{\sqrt{3}}\left(s_{\zeta}-\frac{t^2c_{\zeta}}{\sqrt{2t^2+3}}\right)$ \\
  \hline
\end{tabular}
  \caption{$ZHH$ couplings}\label{Zhh}
\end{table}

  If the momentum of the $Z'$ boson is $p_{\mu}$ then we have $p^2=m^2_{Z'}$ and $p=k_1+k_2$. The amplitude of the decay is
 $$i\mathcal{M}(Z'\rightarrow S_iS_j)=-g^{S_{ij}}_{Z'} \left( k_2-k_1\right).\varepsilon$$
 with $\varepsilon_{\mu}=\varepsilon_{\mu}(p,\lambda_{Z'})$  being the polarization vector of the $Z'$.

 Averaging over the $Z'$ polarization using
 \be  \frac{1}{3}\sum_{\lambda_{Z'}} \varepsilon_{\mu}\varepsilon^*_{\nu}=\frac{1}{3}
 \left( -g_{\mu\nu}+\frac{p_{\mu}p_{\mu}}{m^2_{Z'}}\right)\label{sphancucZp}\ee we obtain the squared amplitude
 \be \frac{1}{3}|\mathcal{M}(Z'\rightarrow S_iS_j)|^2=\frac{1}{3}\left| g^{S_{ij}}_{Z'}
 \right|^2\left[-(k_1-k_2)^2+\frac{\left(k_1^2-k^2_2\right)^2}{m^2_{Z'}}\right]. \label{samplitude1}\ee
  Noting that $ k_1^2=m^2_{S_i},~k_2^2=m^2_{S_j}$ and $(k_1-k_2)^2=2\left( m^2_{S_i}+m^2_{S_j}\right)-m^2_{Z'}$
   we have formula of  $\Gamma(Z'\rightarrow SS)$, namely
  \bea \Gamma(Z'\rightarrow S_iS_j)&=&\frac{1}{16 \pi m_{Z'}}
  \sqrt{\left(1-\frac{m^2_{S_i}+m^2_{S_j}}{m^2_{Z'}}\right)^2-
  \frac{4m^2_{S_i}m^2_{S_j}}{m^4_{Z'}}}\times \frac{1}{3}|\mathcal{M}(Z'\rightarrow S_iS_j)|^2 \crn&=&
   \frac{\left| g^{S_{ij}}_{Z'}\right|^2 m_{Z'} }{48\pi} \times \sqrt{\left(1-\frac{m^2_{S_i}+m^2_{S_j}}{m^2_{Z'}}\right)^2-
   \frac{4m^2_{S_i}m^2_{S_j}}{m^4_{Z'}}} \crn &\times&\left[1-\frac{2(m^2_{S_i}+m^2_{S_j})}{m^2_{Z'}}+
   \frac{\left(m^2_{S_i}-m^2_{S_j}\right)^2}{m^4_{Z'}}\right] \label{zptoss}\eea
  for two distinguishable final states.  For identical final states, there need
  an extra factor $1/2$ to avoid counting each final state twice \cite{martin2}.
Therefore,  if $S_i\equiv S_j\rightarrow S$, then $m_{S_i}
=m_{S_j}=m_S$ and denoting
   $g^{S_{ij}}_{Z'}=g^{S}_{Z'}$ we have more simple formula:
  \be \Gamma(Z'\rightarrow SS)= \frac{| g^{S}_{Z'}|^2m_{Z'}}{96\pi }\left[ 1-
  \frac{4m^2_{S}}{m^2_{Z'}}\right]^{5/2} .\label{zptoss2} \ee
It is noted that $ | g^{S}_{Z'}|^2\sim g^2/12\times \mathcal{O}(1)$, as shown in table
\ref{HHVvertex} for DCHs.  This means that
 \be  \Gamma(Z'\rightarrow SS) \sim \frac{g^2m_{Z'}}{576\pi}\times \mathcal{O}(1)\ll
  \Gamma(Z'\rightarrow ff). \label{zptwoscalar}\ee

   \subsection{Decay of $Z'$ to one gauge boson  and one scalar}
   This case happens only with scalars that inherit non-zero VEVs values,
    i.e,  neutral Higgses in the model. Detailed investigation shows that possible
  vertices  are  $Z'H^{++}U^{--}$,  $Z'H^{--}U^{++}$ and $Z'ZH^{0}$. This part of Lagrangian has form
     $  g^{SV}_{Z'}Z'_{\mu}V^{\mu}S$.  Vertex factors are represented in Table \ref{zphv}.
\begin{table}
  \centering
  \begin{tabular}{|c|c|c|c|}
    \hline
    Vertex & factor & Vertex & factor \\
    \hline
  $Z'U^{\pm\pm}\rho^{\mp\mp}$ & $\frac{igs_{\gamma}m_W}{\sqrt{3}}\left(s_{\zeta}-\frac{2c_{\zeta}t^2}{\sqrt{2t^2+3}}\right)$
   & $Z'U^{\pm\pm}\rho^{\prime\mp\mp}$& $\frac{igc_{\gamma}m_W}{\sqrt{3}}
   \left(s_{\zeta}-\frac{2c_{\zeta}t^2}{\sqrt{2t^2+3}}\right)$ \\
    \hline
          $Z'U^{\pm\pm}\chi^{\mp\mp}$&  $\frac{igs_{\beta}m_V}{\sqrt{3}}\left(s_{\zeta}+\frac{2c_{\zeta}t^2}
          {\sqrt{2t^2+3}}\right)$& $Z'U^{\pm\pm}\chi^{\prime\mp\mp}$ &
     $\frac{igm_{V}c_{\beta}}{\sqrt{3}}\left(s_{\zeta}+\frac{2c_{\zeta}t^2}
          {\sqrt{2t^2+3}}\right)$  \\
    \hline
   $Z'ZH_{\rho}$ & $\frac{2igc_{2\zeta}s_{\gamma}\sqrt{2t^2+3}m_V^2m_W}{3(m^2_V+m^2_W)}$&
    $Z'ZH_{\rho'}$ & $\frac{2igc_{2\zeta}c_{\gamma}\sqrt{2t^2+3}m_V^2 m_W}{3(m^2_V+m^2_W)}$  \\
     \hline
   $Z'ZH_{\chi}$ & $-\frac{2igc_{2\zeta}s_{\beta}\sqrt{2t^2+3}m_V m^2_W}{3(m^2_V+m^2_W)}$&
    $Z'ZH_{\chi'}$ & -$\frac{2igc_{2\zeta}c_{\beta}\sqrt{2t^2+3}m_Vm^2_W}{3(m^2_V+m^2_W)}$  \\
    \hline
  \end{tabular}
  \caption{Coupling  of $Z'HV$}\label{zphv}
\end{table}
    The partial decay width for this case is
   \bea  \Gamma(Z'\rightarrow SV)&=&   \frac{\left| g^{SV}_{Z'}\right|^2}{48\pi m_{Z'}}\times
    \sqrt{\left(1-\frac{m^2_V+m^2_S}{m^2_{Z'}}\right)^2-\frac{4m^2_Vm^2_S}{m^4_{Z'}}} \crn
   &\times&\left[ 2+
   \frac{\left(m^2_{Z'}+m^2_V-m^2_S\right)^2}{4m^2_Vm^2_{Z'}}\right].\label{Zpsvdecay1}\eea
We can estimate the largest contributions to $
\Gamma(Z'\rightarrow SV)$ are from the $\chi$ and $\chi'$, namely
$ \Gamma(Z'\rightarrow SV)= 0.06 g^2 m_{Z'}\times \mathcal{O}(1)$.

  \subsection{Decay of $Z'$ to gauge boson pairs}

  The possible decays are $Z'\rightarrow WW,VV,UU$  with the respective couplings shown in the table
  \ref{gzpvvp}.
   The  general vertex factor  is $i g^{X}_{Z'}\left[ g^{\mu\nu}(p-k_1)^{\sigma}+g^{\sigma\nu}(k_1-k_2)^{\mu}
   +g^{\mu\sigma}(k_2-p)^{\nu}\right]$, where $X=W,~U$ or $V$ gauge bosons. The partial decay for each particle
   can be written by
  \bea \Gamma(Z'\rightarrow XX)&=&   \frac{\left| g^{X}_{Z'}\right|^2m_{Z'}}{192\pi } \left[1- \frac{4m^2_X}
  {m_{Z'}^2}\right]^{3/2}  \frac{m^4_{Z'}+12m^4_{X}+20m^2_{X}m^2_{Z'}}{m^4_X}.\label{Zpsvdecay1}\eea
From the  mass spectrum of gauge bosons given in appendix \ref{apllv}, we can see that in case of
 $m_W \ll  m_V$ we have $ m^2_{Z'}\simeq \frac{2(t^2+2)}{3}  m^2_U$ and $m^2_U\simeq m^2_V$. Furthermore, from the definition
 of $\zeta$ in \ref{dzeta} we can see that in the limit of $m_W/m_V\rightarrow 0$ we get $g^W_{Z'}\rightarrow 0$.
 More exactly if $m_W^2/m_V=\epsilon \ll 1$  then $g^W_{Z'} \simeq\frac{\sqrt{3}g (t^2+1)^2}{\sqrt{2}(t^2+2)^{3/2}}
 \times\frac{m_W^2}{m^2_V}$,  leading to the result that
  $\Gamma(Z'\rightarrow W^+W^-)\simeq \frac{g^2 m_{Z'}(1+t^2)^2}{648\pi(t^2+2)}$.  It is noted that in this case
 $c_{\zeta}\simeq 0.988$ and $s_{\zeta} \simeq 0.155$.

\section{Coupling of doubly charged Higgs}
Three-vertex coupling shown in table \ref{hcc3coupling}
\begin{table}[h]
  \centering
  \begin{tabular}{|c|c|c|c|}
    \hline
    Vertex & factor & Vertex & factor \\
    \hline
   $\rho^{++}\rho^{--} Z^{\mu}$ & $-\frac{ig}{2\sqrt{3}}\left[ \frac{(2t^2-3)
    s_{\zeta}}{\sqrt{2t^2+3}}+c_{\zeta}\right](p+p')_{\mu}$
     & $\rho^{\prime++}\rho^{\prime--} Z^{\mu}$ & $-\frac{i g}{2\sqrt{3}}
    \left[\frac{(2t^2-3)s_{\zeta}}{\sqrt{2t^2+3}}+c_{\zeta}\right](p+p')_{\mu}$  \\
    $\chi^{++}\chi^{--} Z^{\mu}$ & $-\frac{i g}{2\sqrt{3}}
    \left[\frac{(2t^2-3)s_{\zeta}}{\sqrt{2t^2+3}}-c_{\zeta}\right](p+p')_{\mu}$
    &  $\chi^{\prime++}\chi^{\prime--} Z^{\mu}$ & $-\frac{i g}{2\sqrt{3}}
    \left[\frac{(2t^2-3)s_{\zeta}}{\sqrt{2t^2+3}}-c_{\zeta}\right](p+p')_{\mu}$  \\
    $\rho^{++}U^{--\mu}H_{\rho}$ & $\frac{i g}{2}(p+p')_{\mu}$  &
    $\rho^{++}U^{--\mu}H_{A_1}$ & $\frac{gc_{\gamma}}{2}(p+p')_{\mu}$  \\
    $\rho^{++}V^{-\mu}H^{-}_{1}$ &
    $-\frac{igc_{\gamma}}{\sqrt{2}}(p+p')_{\mu}$&
    $\rho^{\prime++}U^{--\mu}H_{\rho'}$ & $-\frac{i g}{2}(p+p')_{\mu}$  \\
    $\rho^{\prime++}U^{--\mu}H_{A_1}$ &
    $\frac{-gs_{\gamma}}{2}(p+p')_{\mu}$&
    $\rho^{\prime++}V^{-\mu}H^{-}_{1}$ & $\frac{i gs_{\gamma}}{\sqrt{2}}(p+p')_{\mu}$  \\
    $\chi^{++}U^{--\mu}H_{\chi}$ & $-\frac{i g}{2}(p+p')_{\mu}$&
    $\chi^{++}U^{--\mu}H_{A_2}$ & $\frac{gc_{\beta}}{2}(p+p')_{\mu}$  \\
    $\chi^{++}W^{-\mu}H^{-}_{2}$ &
    $\frac{i gc_{\beta}}{\sqrt{2}}(p+p')_{\mu}$&
    $\chi^{\prime++}U^{--\mu}H_{\chi'}$ & $-\frac{i g}{2}(p+p')_{\mu}$  \\
    $\chi^{\prime++}U^{--\mu}H_{A_2}$ &
    $-\frac{gs_{\beta}}{2}(p+p')_{\mu}$&
    $\chi^{\prime++}W^{-\mu}H^{-}_{2}$ & $-\frac{ig s_{\beta}}{\sqrt{2}}(p+p')_{\mu}$  \\
    $\rho^{\prime--}l^c_il^c_i$ &
    $\frac{i g m_l}{\sqrt{2}m_W c_{\gamma}c_{\beta}}$&
    $\chi^{\prime++}ll$ &   $\frac{ig m_l}{\sqrt{2}m_V c_{\gamma}c_{\beta}}$ \\

     \hline
  \end{tabular}
  \caption{Three-vertex coupling of doubly charged
  Higgses}\label{hcc3coupling}
\end{table}

Four coupling vertices are listed in table \ref{hcc4coupling}.
\begin{table}[h]
  \centering
  \begin{tabular}{|c|c|c|c|}
    \hline
    Vertex & factor & Vertex & factor \\
    \hline
   $\rho^{++}\rho^{--} A^{\mu}A_{\mu}$ & $4ie^2$
     & $\rho^{\prime++}\rho^{\prime--} A^{\mu}A_{\mu}$ & $4ie^2$ \\
     $\chi^{++}\chi^{--} A^{\mu}A_{\mu}$ & $4ie^2$
     & $\chi^{\prime++}\chi^{\prime--} A^{\mu}A_{\mu}$ & $4ie^2$ \\
     $\rho^{++}\rho^{--} A^{\mu}Z_{\mu}$ & $\frac{-i\sqrt{2}g^2 t}{\sqrt{3(2t^2+3)}}\left[
     \frac{(2t^2-3)s_{\zeta}}{\sqrt{2t^2+3}}+c_{\zeta}\right]$
     & $\rho^{\prime++}\rho^{\prime--} A^{\mu}Z_{\mu}$ & $\frac{-i\sqrt{2}g^2 t}{\sqrt{3(2t^2+3)}}\left[
     \frac{(2t^2-3)s_{\zeta}}{\sqrt{2t^2+3}}+c_{\zeta}\right]$ \\
     $\chi^{++}\chi^{--} A^{\mu}Z_{\mu}$ & $\frac{-i\sqrt{2}g^2 t}{\sqrt{3(2t^2+3)}}\left[
     \frac{(2t^2-3)s_{\zeta}}{\sqrt{2t^2+3}}-c_{\zeta}\right]$
     & $\chi^{\prime++}\chi^{\prime--} A^{\mu}Z_{\mu}$ & $\frac{-i\sqrt{2}g^2 t}{\sqrt{3(2t^2+3)}}\left[
     \frac{(2t^2-3)s_{\zeta}}{\sqrt{2t^2+3}}-c_{\zeta}\right]$ \\
     $ \rho^{++}H_{\rho} U^{--\mu}A_{\mu}$& $\frac{ie^2}{s_W}$  &
     $ \rho^{\prime++}H_{\rho'} U^{--\mu}A_{\mu}$& $\frac{ie^2}{s_W}$ \\
     $ \chi^{++}H_{\chi} U^{--\mu}A_{\mu}$& $\frac{ie^2}{s_W}$  &
     $ \chi^{\prime++}H_{\chi'} U^{--\mu}A_{\mu}$& $\frac{ie^2}{s_W}$ \\
     $ \rho^{++}H_{\rho} U^{--\mu}Z_{\mu}$& $-\frac{ig^2}{2\sqrt{3}}
     \left[\frac{2 t^2 s_{\zeta}}{\sqrt{2t^2+3}}+c_{\zeta}\right]$
     &$ \rho^{\prime++}H_{\rho'} U^{--\mu}Z_{\mu}$& $-\frac{ig^2}{2\sqrt{3}}
     \left[\frac{2 t^2 s_{\zeta}}{\sqrt{2t^2+3}}+c_{\zeta}\right]$\\
     $ \chi^{++}H_{\chi} U^{--\mu}Z_{\mu}$& $\frac{ig^2}{2\sqrt{3}}
     \left[\frac{2 t^2 s_{\zeta}}{\sqrt{2t^2+3}}-c_{\zeta}\right]$
     &$ \chi^{\prime++}H_{\chi'} U^{--\mu}Z_{\mu}$& $\frac{ig^2}{2\sqrt{3}}
     \left[\frac{2 t^2 s_{\zeta}}{\sqrt{2t^2+3}}-c_{\zeta}\right]$\\
     $ \rho^{++}H_{\rho} V^{-\mu}W^{-}_{\mu}$& $\frac{ig^2}{2\sqrt{2}}$
     &$ \rho^{\prime++}H_{\rho'} V^{-\mu}W^{-}_{\mu}$& $\frac{ig^2}{2\sqrt{2}}$\\
     $ \chi^{++}H_{\chi} V^{-\mu}W^{-}_{\mu}$& $\frac{ig^2}{2\sqrt{2}}$
     &$ \chi^{\prime++}H_{\chi'} V^{-\mu}W^{-}_{\mu}$& $\frac{ig^2}{2\sqrt{2}}$\\
     $ \rho^{++}H_{A_1} U^{--\mu}A_{\mu}$& $\frac{e^2c_{\gamma}}{s_W}$  &
     $ \rho^{\prime++}H_{A_1} U^{--\mu}A_{\mu}$& $\frac{-e^2s_{\gamma}}{s_W}$ \\
     $ \chi^{++}H_{A_2} U^{--\mu}A_{\mu}$& $\frac{e^2c_{\beta}}{s_W}$  &
     $ \chi^{\prime++}H_{A_2} U^{--\mu}A_{\mu}$& $-\frac{e^2s_{\beta}}{s_W}$ \\
     $ \rho^{++}H_{A_1} U^{--\mu}Z_{\mu}$& $-\frac{g^2c_{\gamma}}{2\sqrt{3}}
     \left[\frac{2 t^2 s_{\zeta}}{\sqrt{2t^2+3}}+c_{\zeta}\right]$
     &$ \rho^{\prime++}H_{A_1} U^{--\mu}Z_{\mu}$& $\frac{g^2s_{\gamma}}{2\sqrt{3}}
     \left[\frac{2 t^2 s_{\zeta}}{\sqrt{2t^2+3}}+c_{\zeta}\right]$\\
     $ \chi^{++}H_{A_2} U^{--\mu}Z_{\mu}$& $-\frac{g^2c_{\beta}}{2\sqrt{3}}
     \left[\frac{2 t^2 s_{\zeta}}{\sqrt{2t^2+3}}-c_{\zeta}\right]$
     &$ \chi^{\prime++}H_{A_2} U^{--\mu}Z_{\mu}$& $\frac{g^2s_{\beta}}{2\sqrt{3}}
     \left[\frac{2 t^2 s_{\zeta}}{\sqrt{2t^2+3}}-c_{\zeta}\right]$\\
     $ \rho^{++}H_{A_1} V^{-\mu}W^{-}_{\mu}$& $\frac{g^2c_{\gamma}}{2\sqrt{2}}$
     &$ \rho^{\prime++}H_{A_1} V^{-\mu}W^{-}_{\mu}$& $\frac{-g^2s_{\gamma}}{2\sqrt{2}}$\\
     $ \chi^{++}H_{A_2} V^{-\mu}W^{-}_{\mu}$& $\frac{-g^2c_{\beta}}{2\sqrt{2}}$
     &$ \chi^{\prime++}H_{A_2} V^{-\mu}W^{-}_{\mu}$& $\frac{g^2s_{\beta}}{2\sqrt{2}}$\\
     $ \rho^{++}H^-_{1} V^{-\mu}A_{\mu}$& $\frac{-3ie^2c_{\gamma}}{\sqrt{2}s_W}$
     &$ \rho^{\prime++}H^-_{1} V^{-\mu}A_{\mu}$& $\frac{3ie^2s_{\gamma}}{\sqrt{2}s_W}$\\
     $ \chi^{++}H^-_{2} W^{-\mu}A_{\mu}$& $\frac{3ie^2c_{\beta}}{\sqrt{2}s_W}$
     &$ \chi^{\prime++}H^-_{2} W^{-\mu}A_{\mu}$& $\frac{-3ie^2s_{\beta}}{\sqrt{2}s_W}$\\
     $ \rho^{++}H^-_{1} U^{--\mu}W^{+}_{\mu}$& $\frac{-ig^2c_{\gamma}}{2}$
     &$ \rho^{\prime++}H^-_{1} U^{--\mu}W^{+}_{\mu}$& $\frac{ig^2s_{\gamma}}{2}$\\
     $ \chi^{++}H^-_{2} U^{--\mu}V^{+}_{\mu}$& $\frac{-ig^2c_{\beta}}{2}$
     &$ \chi^{\prime++}H^-_{2} U^{--\mu}V^{+}_{\mu}$& $\frac{ig^2s_{\beta}}{2}$\\
     $ \rho^{++}H^-_{1} V^{-\mu}Z_{\mu}$& $\frac{ig^2c_{\gamma}}{2\sqrt{6}}
     \left[\frac{(4 t^2-3) s_{\zeta}}{\sqrt{2t^2+3}}-c_{\zeta}\right]$
     &$ \rho^{\prime++}H^-_{1} V^{-\mu}Z_{\mu}$& $-\frac{ig^2 s_{\gamma}}{2\sqrt{6}}
     \left[\frac{(4 t^2-3) s_{\zeta}}{\sqrt{2t^2+3}}-c_{\zeta}\right]$\\
     $ \chi^{++}H^-_{1}W^{-\mu}Z_{\mu}$& $\frac{ig^2c_{\beta}}{2\sqrt{6}}
     \left[\frac{(4 t^2-3) s_{\zeta}}{\sqrt{2t^2+3}}+c_{\zeta}\right]$
     &$ \chi^{\prime++}H^-_{1} W^{-\mu}Z_{\mu}$& $-\frac{ig^2 s_{\beta}}{2\sqrt{6}}
     \left[\frac{(4 t^2-3) s_{\zeta}}{\sqrt{2t^2+3}}+c_{\zeta}\right]$\\
    \hline
  \end{tabular}
  \caption{Four-vertex coupling of DCHs}\label{hcc4coupling}
\end{table}

\end{document}